\documentstyle[aps,prb]{revtex}
\begin{document}
\draft

\title{Memory Loss and Auger Processes in a Many Body Theory of Charge
Transfer}
\author{A.V. Onufriev and J. B. Marston}
\address{Department of Physics, Brown University, Providence, RI 02912-1843}
\date{December 12, 1995}
\maketitle
\centerline{cond-mat/9512102}

\begin{abstract}
Charge transfer between hyperthermal alkali atoms and metallic
scattering surfaces is an
experimental and theoretical arena for many-body interactions.
To model new facets, we use a generalized time-dependent Newns-Anderson
Hamiltonian which includes electron spin,
multiple atomic orbitals with image shifted levels,
intra-atomic Coulomb repulsion, and resonant exchange.
A variational electronic many-body wave function solves the dynamical problem.
The wave function consists of sectors with either zero or one particle-hole
pair and goes beyond earlier work with the inclusion of amplitudes
for a neutral atom plus an electron-hole pair.  Higher order
sectors with more than one particle-hole pair are suppressed by powers of
$1/N$; hence the wave function ansatz is equivalent to a $1/N$
expansion.  The equations of motion are integrated numerically without
further approximation.  The new solution shows improved loss-of-memory --
the final charge state is independent of the initial one --
in agreement with theoretical and experimental expectations.
Understanding of this phenomenon is deepened through an analysis of
entropy production.  By studying the independent-particle approximation, and by
examining the role played by different sectors of the Hilbert space in
entropy production, we arrive at necessary and sufficient conditions
for loss-of-memory to occur in the many-body solution.
As further tests of the theory, we reproduce the experimentally observed
peak in the excited neutral Li(2p) occupancy at intermediate work
functions starting from different initial conditions.
Next, we include Auger processes
by adding two-body interaction terms to the many-body Hamiltonian.
Several types of Auger processes are considered,
and these are shown to affect the final state occupancies
at low work-functions because phase space increases
rapidly as the work-function is lowered.  Preliminary experimental
evidence for an upturn
in the Li(2p) occupancy at the lowest work-functions thus may be explained by
Auger transitions.  Finally, we comment on the plausibility of observing
a signature of the Kondo resonance in charge transfer experiments.
\end{abstract}

\bigskip

\section{INTRODUCTION}
\label{sec:intro}

Charge-transfer between metallic surfaces and atoms is a quantum mechanical
many-body phenomenon.
Electrons of either spin up or spin down can neutralize a positive ion,
but once one species has transferred to the atom, electrons of the opposing
spin are blocked, at least partially, by the two-body Coulomb repulsion $U$.
In previous work\cite{M1} [I] the time-dependent
Newns-Anderson Hamiltonian was employed as a model
of resonant charge transfer dynamics in the
scattering of alkali atoms off metal surfaces.  The only approximation
made in solving the model was a systematic truncation of the Hilbert space.
This variational approach, pioneered in the static case by Varma and
Yafet\cite{VY}, and in the dynamical problem by Brako and Newns\cite{BN2},
is equivalent to
a systematic $1/N$ expansion, where $N$ is the spin degeneracy of the electrons
which equals two for the physical case of spin up and down.  The model and
its approximate solution have been used by two experimental groups
to describe the interaction of
hyperthermal Li, Na, and K ions with an Alkali/Cu(001) surface\cite{BC} and
Li ions with an Alkali/Al(100) surface\cite{Weare}.
Qualitative agreement has been found between experiment and theory.

In this article we extend the many-body model of [I] by adding Auger processes.
We also improve the approximate solution by including higher-order terms.
One test of the accuracy of the approximation is provided by the
phenomenon of loss-of-memory, which is said to occur
when the final state of a dynamical system is independent of its initial state.
It has been experimentally observed\cite{Hermann}
that the relative proportion of charge
species in a scattered beam of atoms depends only on
parameters such as surface work-function and the outgoing velocity.
Loss-of-memory occurs in the independent-particle
approximation to the many-body Newns-Anderson Hamiltonian\cite{LG} and, as
explained below, should also occur in better approximations which respect the
strong intra-atomic
correlation.  To test loss-of-memory in the approximate solution we
integrate the equations of motion forward in time starting from four different
initial conditions. The calculations show a significant improvement of
loss-of-memory compared to that found in [I] with a more restricted
Hilbert space.  By analyzing loss-of-memory in terms of the increase of
entropy, we find a simple explanation for this improvement.

Loss-of-memory is important for another reason.
The Newns-Anderson model breaks down when the atom is in the strong coupling
region very close to the metal surface because the atomic orbitals, which
in the model are assumed to be orthogonal to the metal states, hybridize
with surface states.  Nevertheless, as long as loss-of-memory occurs,
the Newns-Anderson model will be an accurate description of charge transfer
because the final charge state of the outgoing atom is determined on the
outbound portion of its trajectory, beyond the strong coupling region.
The breakdown in the model close to the surface therefore does
not affect the subsequent physics of charge transfer further out.

The outline of the rest of the paper is as follows.
In Sec. \ref{sec:model} we discuss the generalized Newns-Anderson Hamiltonian
of resonant charge transfer.
The approximate solution of the model is presented in Sec. \ref{sec:soln}.
To the Hilbert space originally considered in [I] we
add new sectors to the many-body wave function at order $1/N$ which correspond
to a neutral atom plus a particle-hole pair and solve the resulting
equations of motions numerically.  We compare the solutions to ones
obtained previously in [I] and find that the new model agrees better
with experiment as there is improved loss-of-memory.  We also comment
on the plausibility of observing the Kondo effect in charge
transfer experiments.  Sec. \ref{sec:loss} of the present work is
devoted to analyzing the origin of loss-of-memory.
We study the relationship between loss-of-memory and
growth in a coarse-grained von Neumann entropy.
For comparison, we also calculate the corresponding
entropy increase in the independent-particle approximation.
Since the Hilbert space is unrestricted in the independent-particle
approximation, the comparison clarifies how the truncation of Hilbert space
affects entropy production.   In Sec. \ref{sec:auger}
we add two-body interaction
terms to the original Hamiltonian which model several types of
Auger processes.  A simple phase-space argument shows
that these couplings are increasingly important at low work-functions.
We demonstrate that Auger processes can explain the
experimentally observed\cite{Behringer}
upturn in the formation of excited Li(2p) atoms at the very lowest
work-functions.  Conclusions are presented in Sec. \ref{sec:conc}.

\section{The generalized Newns-Anderson model}
\label{sec:model}

To model the dynamics of charge transfer, we make several
simplifying assumptions.  We employ the
Newns-Anderson Hamiltonian, ignore radiative
charge transfer processes, and for now consider only resonant charge transfer.
The electrons in the target metal are modeled as zero-temperature
non-interacting spinning fermions, albeit with the renormalized dispersion
of a Landau Fermi liquid.\cite{HKM}  The zero-temperature approximation is
justified, as experiments typically operate at temperatures much less
than other relevant electronic energy scales.  The atom is modeled as a
system with
a finite number of discrete states moving along a fixed classical trajectory
given by $z(t)$ where $z$ is the distance from the atom to the metal surface.
Each of these atomic states couples to the metal electrons when the atom is
close to the metal surface.
Feedback between the electronic degrees of freedom and the trajectory is
ignored in the formulation.  This trajectory approximation
should be adequate as long as the kinetic energy of the ion is much larger
than the electronic energies.

The model is defined by the following generalized time-dependent
Newns-Anderson Hamiltonian:
\begin{eqnarray}
H(t) &=& \sum_a [\epsilon^{(1)}_a(z) \hat P_1  + \epsilon^{(2)}_a(z) \hat P_2]~
\ c_a^{\dag \sigma} c_{a \sigma}
 +  \sum_k \epsilon_k\ c_k^{\dag \sigma} c_{k \sigma}
\nonumber \\
&+& N^{-1/2} \sum_{a;\ k} \{ [V^{(1)}_{a;k}(z) \hat P_1
+ V^{(2)}_{a;k}(z) \hat P_2]
\ c_a^{\dag \sigma} c_{k \sigma} + H.c \}
\nonumber \\
&+&  {{1}\over{2}} ~ \sum_a U_{aa} n_a (n_a - 1)
+ \sum_{a > b} U_{ab} n_a n_b ~.
\label{O.H}
\end{eqnarray}
Here the fermion operator
$c_a^{\dagger \sigma}$ creates a spin $\sigma$ electron in
orbital $a$ of the atom. For Lithium, a = 0 for the 2s orbital, a = 1, 2,
and 3 for
2p$_z$, 2p$_x$, and 2p$_y$, etc.  Likewise, $c_k^{\dagger \sigma}$ creates an
electron of momentum $k$ and energy $\epsilon_k$
in the metal.  Of course, $k$ is really a three-vector which labels all of
the levels in the metal, both filled and empty,
but it may be regarded as a scalar without loss of generality by absorbing
the three-dimensional aspects of the problem into $\epsilon_k$
and $V_{a;k}$.  We introduce the operators $\hat P_1$ and $\hat P_2$ to
project respectively onto
atoms with one or two valence electrons.  These projectors, which may
be written in terms of the orbital occupancies
$n_a \equiv c_a^{\dagger \sigma} c_{a \sigma}$,
permit one to
assign different orbital energies, $\epsilon^{(1)}_a$ and $\epsilon^{(2)}_a$,
and metal-atom couplings, $V^{(1)}_{a;k}$ and $V^{(2)}_{a;k}$,
to the two cases of neutral atoms and negative ions.
An implicit sum over repeated upper and
lower Greek indices is adopted; for now $N = 2$ and $\sigma = 1,\ 2$
to represent the physical SU(2) case of spin up and down electrons.
We have multiplied the atom-metal resonant coupling by a factor of $N^{-1/2}$.
This factor keeps atomic level widths finite in the
$N \rightarrow \infty$ limit.
Finally, we eliminate excited negative ions from the Hilbert space
by taking the Coulomb repulsion $U_{a b} \rightarrow \infty$ for $a, b \neq 0$.

The orbital energies and atom-metal couplings change with time.
Time-dependence enters through the ion trajectory, which we model as:
\begin{eqnarray}
z(t) &=& z_f - u_i * t;\ t \leq t_{turn} \equiv (z_f - z_0)/u_i\ .
\nonumber \\
&=& z_0 + u_f*(t - t_{turn});\ t > t_{turn}.
\label{trajectory}
\end{eqnarray}
Thus the trajectory starts at a distance $z_f$ far away from the surface at
time $t=0$.  We account roughly for the decrease in ion kinetic energy during
impact, due principally to the recoil of surface atoms and the
change in the scattering angle, by instantaneously changing the initial
perpendicular
component of the ion velocity, $u_i$, to $u_f < u_i$ at the point of closest
approach, $z_0$.

The Fermi energy $\epsilon_F$ is defined to be zero and
the vacuum level lies above $\epsilon_F$ at work-function $W$.
For simplicity, we define all orbital energies $\epsilon_a$ relative
to $\epsilon_F$. Because of image charges,
the orbital energies of the neutral atom $\epsilon^{(1)}_a$ shift upward
by $e^2/4z$ as the atom
approaches the metal surface. To parametrize this $z$-dependence we use
the following form for $\epsilon^{(1)}_a$ which saturates close to the surface:
\begin{eqnarray}
{\epsilon^{(1)}_a(z) } &=&  I_a + W +
(1/v_{max}^2 + 16 (z - z_{im})^2 / e^4)^{-1/2}~ , z > z_{im}
\nonumber \\
&=& I_a + W + v_{max}~ , z < z_{im}
\label{ionshift}
\end{eqnarray}
Here $I_a$ is the ionization energy of an orbital $a$ of an
isolated atom which is taken to be negative and
$z_{im}$ is the distance from the surface at which the image shift saturates
to the value $v_{max}$.

In contrast to the ionization levels, the affinity levels shift downward
as the atom approaches the surface.  In other words, the energy required to
remove the two valence electrons bound to a negative alkali ion
(thereby making it a positive ion) is unaffected by the image charges.
As the intra-atomic Coulomb repulsion is already
accounted for explicitly in the two-body interaction
term in Eq. (\ref{O.H}), the orbital energies for the negative ion are given
by the same formula as Eq. (\ref{ionshift}) without the image shift:
\begin{equation}
\epsilon^{(2)}_a = I_a + W \ .
\label{affinshift}
\end{equation}
The Coulomb energy between two electrons in the
lowest s-orbital ($a = 0$) is then given by $U_{00} = A - I_0$ where
$A$ is the electron affinity (also defined here to be negative).
The atom-metal couplings $V_{a;k}$ decay exponentially with
distance when the atom
is far from the metal surface because the atomic wave functions
drop off exponentially with increasing distance from the atom,
and the electronic wave functions in the metal fall off exponentially
with increasing $z$.
Closer in, the couplings deviate from the pure exponential form and saturate.
In the following calculations we ignore the $k$ dependence of
the metal-atom coupling.  This approximation
is justified in so far as most of the resonant electronic
processes occur close to the Fermi surface and the wave vector
dependence of the couplings is smooth.

The metal states are labeled by $2M$
discrete momenta, $M$ above the Fermi energy and $M$ below it.
We set $M = 30$ in the numerical calculations presented below, a
sufficient number to sample the continuum of states accurately.
Though the couplings $V_{a;k}$ are
of fundamental importance in the many-body theory, it is convenient to
express them in terms of the atomic half-widths, as the couplings must
be rescaled each time we change the number of discrete metal states, $M$.
We relate couplings and half-widths $\Delta_a$ via the
approximate independent-particle Fermi Golden Rule formula:
\begin{equation}
V_{a;k}^2 = {{\Delta_a~}\over{\pi \rho}}
\label{FGR}
\end{equation}
where $\rho = M/D$ is the density of states for a flat band of half width $D$.
Level half-widths $\Delta_a(z)$ are obtained from first principle calculations,
within an independent-particle approximation, carried out by
Nordlander and Tully\cite{PN} and Nordlander\cite{PN2}.  Exact values for
$V_{a;k}$ will, of course, differ somewhat from those obtained via
Eq. (\ref{FGR}).  To be useful, theoretical predictions must be robust to
changes in the values of the couplings.  A simple 3-parameter function,
which accounts
both for exponential decrease away from the surface and saturation close to it,
fits the calculated widths well\cite{Ernie}:
\begin{equation}
{\Delta_a(z)} = {{\Delta_0}\over{[{e^{4 \alpha z}}
{}~ +~ {{({\Delta_0}/{\Delta_{sat}})}^4~ -~ 1~ ]}^{1/4}}}\ .
\label{widthfit}
\end{equation}
To be concrete,
we study the case of Lithium atoms interacting with a Cu(001) surface.
Some of the parameters which appear in the Hamiltonian Eq. (\ref{O.H})
via equations Eq. (\ref{trajectory}), Eq. (\ref{ionshift}),
and Eq. (\ref{widthfit}) are fixed throughout the rest of the paper.
In Eq. (\ref{trajectory}) we either start the trajectory far away from the
surface at $z_f = 20 \AA$ and bounce off the surface at $z_0 = 1 \AA$ or
begin from the point of closest approach, $z_0$, and integrate outward.
In Eq. (\ref{ionshift}) we take $z_{im} = 0.0 \AA $ and $v_{max} = 2.6$ eV.
For Lithium, the ionization energy from the 2s ground state is given by
$I_0 = -5.39$ eV and the ionization energy from the 2p$_z$
excited state is $I_1 = -3.54$ eV.
We ignore the 2p$_{x,y}$ states as they couple
only weakly to the metal.  We also eliminate higher lying excited states and,
as mentioned above,
excited states of the negative ion as these states are not expected to become
significantly populated.
The electron affinity energy in Eq. (\ref{affinshift}) is given by
$A = -0.62$ eV.  Finally, the half-bandwidth of Copper is given approximately
by $D = 4$ eV.  Parameters appearing in the resonant widths
formula Eq. (\ref{widthfit}) are given in Table \ref{table1}.
Parameters which vary
are the surface work-function $W$ and the incoming and the outgoing velocities
of the Lithium atom $u_i$ and $u_f$.
Values for these variables are listed in the text below and in
the figure captions.

\section{Systematic Solution}
\label{sec:soln}

To construct an approximate wave function for the problem we follow
Varma and Yafet\cite{VY} and also
Brako and Newns\cite{BN2} and group the full many-body electronic
wave function into sectors containing more and more numbers of
particle-hole excitations in the metal.  Upon truncating the wave function
at a given number of particle-hole pairs, we obtain a variational
wave function that spans only a small, but manageable,
portion of the entire Hilbert space.
The amplitude for particle-hole pair production
is controlled at least formally by generalizing the
two types of SU(2) electrons (spin up and down) to N types of SU(N)
fermions.  Thus the spin index $\sigma$ now runs from 1 to N.
We show below that the amplitudes for terms involving more and
more particle-hole pairs are reduced by higher and higher powers of 1/N.

To begin, we decompose the many-body wave function into five sectors,
four of which were introduced in [I]. The new fifth sector consists of
two parts, symmetric and antisymmetric.  In this paper we adopt the convention
of using capital letters to denote momenta indices which are restricted to
values greater than $k_F$, or in other words, states above the Fermi energy.
Lower case letters denote momenta indices which run over values less than
$k_F$.  The variational ansatz for the many-body wavefunction can then be
written as:
\begin{eqnarray}
| \Psi(t) \rangle &=& f(t) | 0 \rangle +
\sum_{a;\ k<k_F} b_{a;k}(t) |a; k \rangle
+ \sum_{k<k_F,\ L>k_F} e_{L,k}(t) |L, k \rangle
+ \sum_{q<k<k_F} d_{k,q}(t) |k, q \rangle
\nonumber \\
&+& \underbrace{\sum_{a; L>k_F,\ q<k<k_F} s_{a; L,k,q}(t)
{|a; L, k, q \rangle}^{S}
+ \sum_{a; L>k_F,\ q<k<k_F} a_{a; L,k,q}(t) {|a; L, k, q \rangle}^{A}}
_{\bf new~sectors}
\nonumber \\
&+& (rest\ of\ Hilbert\ space)\ .
\label{O.PSI}
\end{eqnarray}
Each sector is a global SU(N) singlet.  Non-singlet sectors can be ignored
in so far as the initial state of the system,
a closed shell positive alkali ion far away from an unperturbed non-magnetic
metal, and the Hamiltonian are both SU(N) singlets.  Here the orthonormal
basis states in different sectors of the Hilbert space are given by:
\begin{eqnarray}
|a; k \rangle &\equiv& N^{-1/2}\ c_a^{\dagger
\sigma} c_{k \sigma} | 0 \rangle\ .
\nonumber \\
|L, k \rangle &\equiv& N^{-1/2}\ c_L^{\dagger \sigma} c_{k \sigma}
| 0 \rangle \ .
\nonumber \\
|k, q \rangle
&\equiv& [N(N-1)]^{-1/2}\ c_{0}^{\dagger \alpha}
c_{k \alpha}
c_{0}^{\dagger \beta}
c_{q \beta} | 0 \rangle\ .
\label{O.Sectors}
\end{eqnarray}
The basis for the new sectors is given by:
\begin{eqnarray}
{|a; L, k, q \rangle}^{S}
&\equiv& [2N(N-1)]^{-1/2}\lbrace\ c_{L}^{\dagger \alpha}
c_{k \alpha} c_{a}^{\dagger \beta} c_{q \beta} | 0 \rangle\
+ \ c_{L}^{\dagger \alpha} c_{q \alpha} c_{a}^{\dagger \beta}
c_{k \beta} | 0 \rangle\ \rbrace.
\nonumber\\
{|a; L, k, q \rangle}^{A}
&\equiv& [2N(N+1)]^{-1/2}\lbrace\ c_{L}^{\dagger \alpha}
c_{k \alpha} c_{a}^{\dagger \beta} c_{q \beta} | 0 \rangle\
- \ c_{L}^{\dagger \alpha} c_{q \alpha} c_{a}^{\dagger \beta}
c_{k \beta} | 0 \rangle\ \rbrace.
\label{N.Sector}
\end{eqnarray}
The reference state $| 0 \rangle$ represents a positive alkali ion
(i.e. an empty valence shell) along with the Landau
Fermi-liquid at zero-temperature with no particle-hole
excitations.  According to the convention the limits on the momenta ranges
appearing in Eqs. (\ref{O.Sectors}) and (\ref{N.Sector})
are shorthand notation for $\epsilon_q < \epsilon_k
< \epsilon_F$ and $\epsilon_L > \epsilon_F$
where $\epsilon_F \equiv 0$ is the Fermi energy.
In other words, $k$ and $q$ label
hole momenta, and $L$ labels particle momentum, so while
$|L, q \rangle$ is a positive ion plus a particle-hole pair, the
state $|k, q \rangle$ instead represents a negative ion with
two holes in the metal.  A schematic of the different sectors of the Hilbert
space is presented in Fig. \ref{fig1}.  We show below that terms involving
two or more particle-hole pairs constitute higher-order corrections which
are dropped in the approximate solution.

The time-dependent coefficients appearing in the many-body wave function
Eq. (\ref{O.PSI}) are amplitudes for the following states:

\indent{(1) $f(t)$ ---
A positive ion with no excitations in the metal, which is at absolute zero
temperature.  Note that
$f(t=0) = 1$ describes the initial state of
an experiment which directs incoming positive ions against the metal
target.}

\indent{(2) $b_{a;k}(t)$ --- A neutral atom with orbital $a$ occupied
and a hole left behind in the metal at momentum $k$.}

\indent{(3) $e_{L,q}(t)$ --- A positive ion and a single
particle-hole pair in the metal (the electron has momentum $L$
and the hole has momentum $q$).}

\indent{(4) $d_{k,q}(t)$ ---
A negative ion with a double-occupied s-orbital ($a = 0$)
and two holes in the metal at momenta $k$ and $q$.}

\indent{(5) $s_{a; L,k,q}$ and $a_{a; L,k,q}$ ---
Amplitudes for the new states which
represent a neutral atom with orbital $a$ occupied plus two holes
in the metal with momenta $k$ and $q$ and a particle of momentum $L$.
To enforce orthogonality,
the sector is split into symmetric ($s$) and antisymmetric parts ($a$)
with respect to interchange of momenta indices $k$ and $q$.  Physically,
the state produced by an electron hopping to the atom from a metallic
level $k$ while another electron hops from $q$ to $L$ can be distinguished
from the state in which $k$ and $q$ are interchanged.
In the special case of no spin degeneracy $N=1$, however,
there is only one state, the antisymmetric one,
as the particles are then spinless and can no longer be distinguished.

The logic behind the truncation scheme
becomes clear upon considering the nature of the off-diagonal
coupling, the terms in the Hamiltonian proportional to $N^{-1/2}~ V_{a;k}$.
These terms
couple adjacent sectors of the Hilbert space, as shown in Fig. \ref{fig2}.
(By adjacent we mean sectors
that differ by at most one elementary excitation in the band like a hole
or a particle.)  Repeated
applications of the off-diagonal coupling to the reference state $|0\rangle$
generates all of the sectors in the singlet many-body wave function.
Each time $V_{a;k}$ acts, it brings along a factor of $N^{-1/2}$.  Thus
amplitudes for sectors involving multiple particle-hole pairs are weakly
coupled to lower order terms when N is large.  In particular, from
Eq. (\ref{EOM}) below it is clear
that the amplitudes of sectors containing a single particle-hole pair
($e_{L, q}$, $s_{a; L,k,q}$ and $a_{a; L, k, q}$) are reduced by a factor
of $N^{-1/2}$ in comparison to the amplitudes for the sectors with
no particle-hole pairs ($f$, $b_{a;k}$
and $d_{k,q}$).  The probability for a particle-hole pair is
therefore reduced by a factor of $1/N$.
The restriction to this trial basis is achieved
by projecting the Schr\"odinger equation
$i {d\over{dt}} \Psi = {\hat H} \Psi$ onto each sector of the Hilbert space
to obtain the equations of motion. Following [I], to reduce computational work
we remove diagonal terms from the equations of motion
by a change of variables:
$\lambda(t) = \Lambda(t) {\exp}^{-i\phi(t)}$ where $\lambda(t)$
is an amplitude and $\phi(t)$ would be the phase of the corresponding state
were the coupling of the atom to the surface turned off.
For instance, in the newly added sector, diagonalization is accomplished
with the following change of variables:
\begin{eqnarray*}
s_{a; L,k,q}(t) & = & S_{a; L,k,q}(t)\
{\exp}\{-i[ \phi_a(t) + ( \epsilon_L - \epsilon_k - \epsilon_q ) t] \}
\nonumber\\
a_{a; L,k,q}(t) & = & A_{a; L,k,q}(t)\
{\exp}\{-i[ \phi_a(t) + ( \epsilon_L - \epsilon_k - \epsilon_q ) t] \}
\end{eqnarray*}
where
$\phi_a(t) \equiv {\int}^{t}_{0} \epsilon^{(1)}_a(t^\prime) dt^\prime$
is the time-evolved
phase for the decoupled, but image-shifted, atomic orbital $a$.
The resulting equations of motion are:
\begin{eqnarray}
i{d\over{dt}}F &=& \sum_{a;\ k} V^{(1)*}_{a;k}
{\exp}\{i[\epsilon_k t - \phi_a(t)]\}\ B_{a;k}\ .
\nonumber\\
i{d\over{dt}}B_{a;k} &=& V^{(1)}_{a;k}\
 {\exp}\{i[\phi_a(t) - \epsilon_k t]\}\ F
\nonumber \\
&+& \delta_{a,0}~ \sqrt{1 - 1/N} \sum_{q} V^{(2)*}_{0;q}\
{\exp}\{-i[(U - \epsilon_q + 2 \epsilon^{(2)}_0) t - \phi_0(t)]\}
[\theta(k-q)~ D_{kq} + \theta(q-k)~ D_{qk}]
\nonumber\\
&+& N^{-1/2} \sum_{L} V^{(1)}_{a;L}\ {\exp}\{i[\phi_a(t) -
\epsilon_Lt]\}\ E_{Lk}\ .
\nonumber\\
i{d\over{dt}}E_{Lk} &=& N^{-1/2} \sum_a V^{(1)*}_{a;L}\ {\exp}
\{i[\epsilon_Lt - \phi_a(t)]\}\ B_{a;k}
\nonumber\\
&+& \sqrt{(N - 1)/2N} \sum_{a;q}V^{(1)*}_{a;q}{\exp}\{i[\epsilon_q t -
\phi_a(t)]\}
[\theta(k-q)\ S_{a;Lkq} + \theta(q-k)\ S_{a;Lqk}]
\nonumber\\
&+& \sqrt{(N + 1)/2N} \sum_{a;q}V^{(1)*}_{a;q}{\exp}\{i[\epsilon_q t -
\phi_a(t)]\}
[\theta(k-q)\ A_{a;Lkq} - \theta(q-k)\ A_{a;Lqk}]\ .
\nonumber\\
i{d\over{dt}}D_{kq} &=& \sqrt{1 - 1/N}
\ V^{(2)}_{0;q}\
{\exp}\{i[(U - \epsilon_q + 2 \epsilon^{(2)}_0) t - \phi_0(t)]\}
\ B_{0;k}
\nonumber\\
&+& \sqrt{1 - 1/N}~ V^{(2)}_{0;k}~
{\exp}\{i[(U - \epsilon_k + 2 \epsilon^{(2)}_0) t - \phi_0(t)]\}~ B_{0;k}
\nonumber\\
&+& (2/N)^{1/2} \sum_{L} V^{(2)}_{0;L}\
{\exp}\{i[(U - \epsilon_L+ 2 \epsilon^{(2)}_0) t - \phi_0(t)]\}~ S_{0;Lkq}\ .
\nonumber\\
i{d\over{dt}}S_{a;Lkq} &=& \delta_{a,0}\ (2/N)^{1/2}\ V^{(2)*}_{0;L}\
{\exp}\{-i[(U - \epsilon_L+ 2 \epsilon^{(2)}_0) t - \phi_0(t)]\}D_{kq}
\nonumber\\
&+& \sqrt{(N - 1)/2N}~ [V^{(1)}_{a;q}{\exp}\{i[\phi_a(t) - \epsilon_q t]\}\
E_{Lk} + V^{(1)}_{a;k}{\exp}\{i[\phi_a(t) - \epsilon_k t]\}\ E_{Lq}]\ .
\nonumber\\
i{d\over{dt}}A_{a; Lkq} &=& \sqrt{(N + 1)/2N}~
[V^{(1)}_{a;q}{\exp}\{i[\phi_a(t) - \epsilon_q t]\}\ E_{Lk}
- V^{(1)}_{a;k}{\exp}\{i[\phi_a(t) - \epsilon_k t]\}\ E_{Lq}]\ .
\label{EOM}
\end{eqnarray}
In the above equations we have used the following symmetries of the amplitudes:
$D_{kq} = D_{qk}$, $S_{a; Lkq} = S_{a; Lqk}$, and
$A_{a; Lkq} = - A_{a; Lqk}$.  Also, we have
corrected several typographical errors which appeared in Eq. (3.6) of [I].
As the amplitudes undergo unitary evolution forward in time, the sum of
their squares is conserved and equals one.

The equations of motion are numerically integrated forward in time
with the use of a fourth-order Runge-Kutta algorithm with adaptive time
steps.  The double-precision C-code is run on IBM RS/6000 machines and, in a
multiprocessor form, on a Cray EL-98 computer.\cite{Brown}
Probability is conserved to better than one part in
$10^6$.  For $M = 30$ levels, one run at a typical velocity takes on the
order of ten minutes of RS/6000 CPU time.
We choose one of the following four initial conditions:

\indent{(1) -- A positive alkali ion A$^+$ far away ($z = z_f$) from
the surface.  The only nonzero initial amplitude is $F(t=0) = 1$.}

\indent{(2) -- A neutral, unexcited, alkali atom A$^0$ far away.
The only nonzero initial amplitude is $B_{0; 0} = 1$. A single hole lies
at the Fermi energy.}

\indent{(3) -- A negative alkali ion A$^-$
far away.  The only nonzero initial amplitude is $D_{0,0} = 1$. Two holes
lie at the Fermi energy.}

\indent{(4) -- Start at the point of closest approach, $z = z_0$,
in the equilibrium ground state. (The ground state
is obtained via the imaginary-time Lanczos algorithm.) This initial
condition is realized in sputtering experiments.}

In Fig. 4(a) we present results from the improved equations of motion
for the case of a Lithium atom striking a clean Cu(001) surface of
work-function $W = 4.59$ eV for three different initial conditions, (1), (2),
and (4), and over a range of velocities ($0.005$ a.u. $< u_f < 0.05$ a.u.).
The occupancies change by less than $1\%$ when the number of metal levels
below the Fermi energy, $M$, is increased from $30$ to $60$.
For comparison, in Fig. 4(b) we also report results obtained from the
previous equations of motion of [I] which are missing the new sectors.
Note in particular the significant improvement in loss-of-memory
compared to that found in [I] for all three initial conditions.  From both
experiment\cite{Hermann,Ernie} and the independent-particle
approximation\cite{BN1}, we expect
loss-of-memory to be complete at this velocity.  Evidently the systematic
$1/N$ expansion works better and better as the Hilbert space is enlarged and
higher order terms are included.  However, we also find that loss-of-memory
is absent from both solutions for initial condition (3), the negative ion.
For this initial condition, the final charge state is nearly 100\% neutral
(A$^0$) for the improved equations of motion, and 100\% negative (A$^-$)
in the case of [I].
The breakdown of loss-of-memory for the negative ion initial condition
has an explanation in the particular manner in which the Hilbert space is
truncated, and we return to this question below in Sec. \ref{sec:loss}.

Another important test of the improved approximation is whether it
reproduces the peak in the excited neutral
Li(2p) occupancy seen in experiments\cite{M1,Behringer}
at a surface work-function value of $W \approx 2.8$ eV.
The improved calculations, like those reported previously in [I], do in
fact yield a peak.  (The physical origin of this peak is discussed in
Sec. \ref{sec:auger} below.)
Experimental measurements\cite{Behringer} of the number of photons produced
by the decay of the excited Li(2p) state to Li(2s) are
plotted alongside the calculated final Li(2p) occupancy in Fig. \ref{fig3}
for the case of initial conditions (2) and (4).  Good qualitative agreement
between theory and experiment is obtained.
The positive ion initial condition does not, however, yield results which
agree with experiment at the work-functions below $2.8$ eV
as the Li(2p) occupancy continues to grow monotonically.
We attribute the breakdown
at low work-functions to the truncation of the Hilbert space.  A term at order
$1/N$ has been left out because it has four momenta indices: the amplitude
in the A$^-$ sector corresponding to a negative ion with
two holes plus a particle-hole pair in the metal.  We expect that
this term could absorb the excess excited neutrals at low work-functions.

Scattering experiments off clean surfaces are the best arena for answering
quantitative questions about
many-body effects, as complications involving surface adsorbates are
then avoided.  Recently Shao {\it et al.} have made an interesting suggestion:
at low velocities, a Kondo resonance should appear in the atomic spectral
function.\cite{Shao2} To argue that this resonance could be observable in
charge-transfer experiments, Shao {\it et al.}
employ a slave boson non-crossing
approximation (NCA) treatment of the time-dependent Newns-Anderson
model\cite{Shao}.  Shao {\it et al.} argue,
based on an assumed pure exponential form for the atom-metal coupling, that
the Kondo peak could show up as deviations from a single straight line
in a plot of the logarithm of the atomic occupancy versus the inverse
perpendicular velocity.

It is important to note that the slave boson NCA approximation breaks down when
the atom is in a mixed valence state, which must occur whenever there is a
level crossing the Fermi energy.  In particular,
a nonphysical temperature scale appears within NCA.
It is an artifact of the approximations made: especially, the neglect of vertex
corrections.\cite{Hewson}  Away from a level crossing it
is small and can usually be neglected.  This is no longer the case in charge
transfer experiments, where the Kondo temperature is of order
$\sqrt{D \Delta}$ near the level crossing.  Then the unphysical temperature
scale is of order $D^2/\Delta$ which is generally larger than both the
bandwidth, $D$, and the Kondo temperature\cite{Hewson}.
A comparison between the NCA
approximation and an essentially exact renormalization group (RG) analysis
of the Anderson impurity model also shows that NCA is an inaccurate
approximation
for dynamical properties.\cite{Costi}  In contrast the $1/N$ approximation
employed here is free of these difficulties as it contains vertex corrections.
For example, the Kondo peak disappears in the variational $1/N$ approach
at $N = 1$
as it should, since there is no longer any spin degeneracy.  In the slave boson
NCA approximation it persists as an unphysical feature.\cite{Hewson}

Besides the technical limitation of the slave boson NCA approximation,
other problems arise in attempting to extract the weak Kondo
signal from the large background.  Shao {\it et al.} assume
that the negative ion has the same width as the
neutral\cite{Shao2} but this assumption should be relaxed since,
as noted above, negative alkali ions
are larger than neutral atoms.  This means negative ion yields cannot be
directly compared to positive yields.  Furthermore,
the detailed form of the width is not a pure exponential and this may
introduce additional non-linearities which will be difficult to separate
from those produced by the Kondo resonance.
Excited states also have been neglected in the model of Shao {\it et al.},
but these may cause wiggles in the occupancy which could be misinterpreted
as Kondo effects.  Indeed, we find no clear signature of the Kondo resonance
in the approximate solution to our model, which incorporates these
generalizations.
Finally, there appears to be no way to do a control experiment in which
only the intra-atomic Coulomb repulsion is turned off, with all else left
unchanged.
Nevertheless, the observation that Kondo effects can in principle occur
serves to underscore the many-body nature of charge-transfer.

\section{Origin of Memory Loss }
\label{sec:loss}
In this section we analyze how the 1/N expansion works in the dynamical
problem.  In particular, we investigate the physical mechanism responsible
for loss-of-memory.  We analyze how loss-of-memory is
affected by the truncation of the Hilbert space to clarify
why the approximate solution exhibits loss-of-memory
for three of the four initial conditions while it breaks down for the case of
an incoming negative alkali ion.  We begin by formulating a simple necessary
criterion for memory loss
to occur and show that it is always satisfied as
long as the initial velocity of the atom is low enough.
As loss-of-memory is not complete in the approximate solution, we conclude
that the conditions which determine its presence or absence are more subtle.
To characterize the
phase decoherence of the initial state and thus loss-of-memory,
we introduce a coarse-grained entropy in
both the independent-particle and in the
many-body pictures.  In the former case
the Hilbert space is unrestricted and loss-of-memory at low
velocities is complete\cite{BN3}.
By comparing entropy increase in the two pictures
we gain insight into the importance of the higher-order sectors left out
in the truncated Hilbert space of Eq. (\ref{O.PSI}).
We show that the probability flow between different
sectors of the Hilbert space is toward the direction of increasing
entropy; the entropy grows as probability flows to sectors which occupy larger
and larger portions of phase space.

First we review the phenomenon of
loss-of-memory within the independent particle approximation.
In this approximation, we neglect the strong correlations
between electrons on the atom by reducing
the atomic states to a single orbital, and by treating the electrons as
spinless ($N = 1$).  Then the Pauli exclusion principle,
instead of the intra-atomic Coulomb
repulsion, prevents multiple occupancy of the atomic orbital.
Consider an atom initially in some state ${|a \rangle}$ incident on the
metal surface.  As the atom moves towards the
surface, it begins to forget its initial state.
If the atom does not spend enough time close to the surface, however,
the initial state will not decay completely, and the final charge state
will depend on the initial one.  Thus the atom must move slowly enough
for loss-of-memory to occur.
In the independent-particle approximation, for the case of infinite bandwidth
($D \rightarrow \infty$), the following expression is obtained for
the time-evolution of the expected atomic occupancy\cite{BN3}:
\begin{equation}
n_{a}(t) = n_{a}(0) \exp \left( -2 \int_{0}^{t}
{\Delta}[z(t^{\prime})] dt^{\prime} \right) + O(t) \ .
\label{occupancy(t)}
\end{equation}
The first term is the memory term $n_a^{mem}$
which depends on the initial atomic occupancy
$n_a(0)$.  The second term $O(t)$
does not interest us here as it is independent of the initial condition.
Assuming pure exponential dependence of the level width on distance,
$\Delta(z) \equiv \Delta_0 \exp(- \alpha z)$, and using
the trajectory approximation Eq. (\ref{trajectory}), the
memory term in (\ref{occupancy(t)}) may be rewritten along the inward bound
portion of the trajectory as:
\begin{equation}
n_{a}^{mem}(t) = n_{a}(0)
\exp \left( - {{2 \Delta_0} \over {\alpha u_{i}}}
\exp[- \alpha (z_f - u_i t)] \right)\ .
\label{memoryterm}
\end{equation}
Initially the atom is at a distance $z_f$ from the surface and moves towards
it with a velocity $u_i$.  It reaches the surface at
$t = t_{turn} = z_f/u_i$ and loss-of-memory is thus complete if
$n_{a}^{mem}(t_{turn}) \ll  n_{a}(0)$ or in other words,
\begin{equation}
2 \Delta_0 >> \alpha u_{i} \ .
\label{condition}
\end{equation}
The physical meaning of this equation is that
there must be enough time for an electron to
hop back and forth between the atom and the metal several times for
loss-of-memory to be complete.
Parameters for the half-widths of Lithium are given in Table {\ref{table1}}.
For the 2s orbital, $\Delta_0 = 2.23$ and
$\alpha = 0.86$.  Thus at a typical incoming velocity of
$u_i = 0.04$ a.u, Eq. (\ref{condition}) is well satisfied.
Although the above estimate is based on the independent-particle approximation
and assumes a pure exponential form for $\Delta(z)$, the conclusion is
valid in the many-body case and for the more general form for $\Delta(z)$ we
use below, as the
key physical feature, the coupling of the atomic level to a continuum of
states in the metal, is unchanged.  Indeed, loss-of-memory occurs within the
slave boson NCA approximate solution to the dynamical many-body problem
in which a class of bubble diagrams is summed
to all orders.\cite{Shao}  In a low-velocity limit the
slave boson NCA approximation reduces to a set of first-order rate
equations which necessarily exhibit loss-of-memory whenever the occupancy
of any channel attains unity along the trajectory.  However,
as quantum mechanical
phase information is thrown away in the semiclassical rate equations,
they are inaccurate at velocities of most experimental interest\cite{BN1,Dahl}
and we do not consider them further here.

It is useful to examine
the time evolution of an atom held at a fixed position close to
the surface as the Hamiltonian is then time-independent.
If there is no loss-of-memory for different initial conditions,
then loss-of-memory will, in general, be absent in the dynamical problem.
Results for the approximate solution to the many-body problem
are presented in Fig. \ref{fig5}.  It typically
takes $\tau_{i} \approx 10^{-15}$ sec for the occupancies to settle down
to constant values.  This interaction
time scale is shorter than the typical amount of time the
atom spends in the region of strong coupling in the dynamical problem,
$\tau_{m} > {5\AA /0.05 a.u.} \approx 5\times 10^{-15}$ sec.
Small oscillations in the occupancies with period $1.03 \times 10^{-15}$
sec at large time are due to the finite metal bandwidth, $D = 4 eV$,
and are washed out in the dynamical system.  The occupancies change by less
than $0.1\%$ when the number of metal states above or below the Fermi energy,
$M$, is increased from 30 to 60.
Note that Poincar{\'e} recurrence, relevant when the coupling between
the atom and the surface is weak, occurs at
the longer time $\tau_{r} = 2 \pi \hbar M / D \approx 2\times 10^{-14}$
sec and can be
ignored.  The final atomic occupancies of about 80\% positive fraction
are nearly the same for
three of the initial conditions - positive ion (1), neutral atom (2),
and equilibrium ground state (4).  In the case of the negative ion initial
condition (3), however, the final charge state is mostly neutral, not positive.

Study of the effects of the truncation of the Hilbert space on
loss-of-memory in the many-body solution
requires a quantitative measure of decoherence.
For this purpose
we may introduce the fine-grained quantum mechanical von Neumann entropy,
$S_{fg}(t) \equiv -Tr\{\hat{\rho}(t) \ln \hat{\rho}(t)\}$,
where $\hat{\rho}(t)$ is the density matrix.
The fine-grained entropy, however, is constant
for any time-independent Hamiltonian
as no phase information is lost in a system undergoing unitary time
evolution.  Thus,
${\rho(t)} = \hat{U}(t) \hat{\rho}(0) \hat{U}^{\dag}(t)$
where $\hat{U}(t) = \exp (-i\hat{H} t)$,
so $S_{fg}(t)= -Tr \{ \hat{U}(t) \hat{\rho}(0) \hat{U}^{\dag}(t)
\hat{U}(t) \ln \hat{\rho}(0) \hat{U}^{\dag}(t) \}$.  Grouping together
$\hat{U}(t)$ and $\hat{U}^{\dag}(t)$ under the $Tr$ symbol, it is
straightforward to see that $S_{fg}(t) =
-Tr\{ \hat{\rho}(0) \ln \hat{\rho}(0) \} = S_{fg}(0)$.
Instead, we coarse-grain\cite{Davies} the system by ignoring information
contained in the off-diagonal matrix elements of $\hat{\rho}$.  The
coarse-grained entropy is then defined to be:
\begin{equation}
S_{cg}(t) = -\sum_{a}{\rho}_{aa}(t) \ln {\rho}_{aa}(t)
\label{EntrDiag}
\end{equation}
where ${\rho}_{aa}$ are the diagonal matrix elements of ${\hat\rho}$
which time evolve as:
\begin{equation}
{\rho}_{aa}(t)=\sum_{b}{|{U_{ab}(t)}|}^2 {\rho}_{bb}(0) + \sum_{b \not= c}
{U_{ab}^{\star}(t)}{U_{ac}(t)~{\rho}_{bc}(0)}\ .
\end{equation}
The second term in this equation contains all the information about phase
correlations, and we expect it to vanish in the $t\rightarrow \infty$
limit provided the Hilbert space is large enough.
Then it is easy to show\cite{Davies} that
$dS_{cg}(t) / dt \geq 0$, the quantum mechanical analogue of the
Boltzmann H-theorem.

It is straightforward to compute the coarse-grained entropy
Eq. (\ref{EntrDiag}) from the many-body states Eq. (\ref{O.PSI}):
\begin{eqnarray}
S_{cg}(t) &=& -{|F|}^2\ln{|F|}^2  - \sum_{a;k} {|B_{ak}|}^2\ln {|B_{ak}|}^2
- \sum_{L,k} {|E_{Lk}|}^2 \ln {|E_{Lk}|}^2
- \sum_{k>q} {|D_{kq}|}^2 \ln {|D_{kq}|}^2
\nonumber \\
&-& \sum_{a;L,k>q} {|S_{a; Lkq}|}^2 \ln {|S_{a; Lkq}|}^2 -
\sum_{a;L,k>q} {|A_{a; Lkq}|}^2 \ln {|A_{a; Lkq}|}^2 - ...
\label{EntrMB}
\end{eqnarray}
where the ellipses denote contributions from higher order sectors not included
in the variational wavefunction.
In the independent-particle picture the coarse-grained entropy is
given by the standard expression for the statistical mechanical
entropy\cite{Landau}:
\begin{equation}
S_{cg}(t) = - \sum_k n_k \ln n_k - \sum_k (1 - n_k) \ln(1 - n_k) - n_a \ln n_a
- (1 - n_a) \ln (1 - n_a)\ .
\label{EntrSP}
\end{equation}
Here $n_k$ is the occupancy of the metal band level $k$ and $n_a$ is the
atomic occupancy.  It is important to note that the two entropy
definitions, Eqs. (\ref{EntrMB}) and (\ref{EntrSP}), are not exactly
equivalent, even for the case of spinless electrons.
Information in the form of two-body and higher order correlations contained
in Eq. (\ref{EntrMB}) has been thrown away in Eq. (\ref{EntrSP})
where only the one-body occupancies appear.  For example, each
Hilbert space sector of Eqs. (\ref{O.Sectors}) and (\ref{N.Sector})
strictly conserves
total particle number.  Conservation of total particle number is reflected in
non-trivial two-body correlations which are discarded when the state is
described purely in terms of one-body occupancies.  While the two entropies
are equal in the limit of a macroscopic number of excitations, for the
finite number of excitations generated in an atom-surface collision the
coarse-grained entropy of Eq. (\ref{EntrSP}) is somewhat
larger than that of Eq. (\ref{EntrMB}).

Occupancies in the independent-particle approximation are obtained by solving
equations of motion for the {\it operators}
$\hat{c}_a(t)$ and $\hat{c}_k(t)$
as opposed to equations of motion for {\it amplitudes} like Eq. (\ref{EOM}).
To highlight this difference, we place hats on top of the operators.  As there
is no many-body interaction U in the independent-particle approximation, the
Heisenberg equations of motion for the operators, obtained from
Eq. (\ref{O.H}), are linear:
\begin{eqnarray}
i {{d}\over{dt}} \hat{c}_a &=& \epsilon^{(1)}_a(t)~ \hat{c}_a
+ \sum_k V^{(1)}_{0;k}(t)~ \hat{c}_k\ ,
\nonumber \\
i {{d}\over{dt}} \hat{c}_k &=& \epsilon_k~ \hat{c}_k + V^{(1)}_{0;k}(t)~
\hat{c}_a \ .
\label{speom}
\end{eqnarray}
Here and below $a = 0$ and the momentum index $k$ runs over all momenta,
not just $k < k_F$.  The
operators at time $t$ may be expressed as a linear combination of the operators
at the initial time $t=0$:
\begin{eqnarray}
\hat{c}_a(t) &=& f(t)~ \hat{c}_a(0) + \sum_k b_k(t)~ \hat{c}_k(0) \ ,
\nonumber \\
\hat{c}_k(t) &=& d_k(t)~ \hat{c}_a(0) + \sum_q e_{kq}(t)~ \hat{c}_q(0) \ .
\label{lincomb}
\end{eqnarray}
Initially, the time-dependent c-number coefficients are given by
$f(0) = 1$, $b_k(0) = d_k(0) = 0$, and $e_{kq}(0) = \delta_{kq}$.
Subsequent values are obtained from the equations of motion for the
coefficients:
\begin{eqnarray}
i {{d}\over{dt}} f &=& \epsilon^{(1)}_a(t)~ f
+ \sum_k V^{(1)}_{0;k}(t)~ d_k\ ,
\nonumber \\
i {{d}\over{dt}} b_k &=& \epsilon^{(1)}_a(t)~ b_k + \sum_q V^{(1)}_{0;q}(t)~
e_{qk} \ ,
\nonumber \\
i {{d}\over{dt}} d_k &=& \epsilon_k~ d_k + V^{(1)}_{0;k}(t)~ f\ ,
\nonumber \\
i {{d}\over{dt}} e_{kq} &=& \epsilon_k~ e_{kq} + V^{(1)}_{0;k}(t)~ b_q \ .
\label{CEOM}
\end{eqnarray}
Once the diagonal terms are removed by a change of variables as in
Eq. (\ref{EOM}), these equations are numerically integrated forward in time
with the fourth-order Runge-Kutta algorithm.\cite{Brown}
The occupancies $n_a(t) = \langle \hat{c}^\dagger_a(t) \hat{c}_a(t) \rangle$
and $n_k(t) = \langle \hat{c}^\dagger_k(t) \hat{c}_k(t) \rangle$
may then be calculated for
any initial state of the system.
For example, in the case of an incident positive alkali ion and a filled
Fermi sea, initially $n_a(0) = 0$ and $n_k(0) = 1$ for $k < k_F$.
The many-electron wavefunction is then
given at all times by the Slater determinant
\begin{equation}
| \Psi(t) \rangle = \Pi_{k<k_F} \hat{c}^\dagger_k(t)~ |0 \rangle = \Pi_{k<k_F}
[d^*_k(t)~ \hat{c}^\dagger_a(0) + \sum_q e^*_{kq}(t)~ \hat{c}^\dagger_q(0)]~
|0 \rangle\ ;
\label{Slater}
\end{equation}
here $|0\rangle$ is the true vacuum state devoid of any electrons.
Now it is clear how an arbitrary number of particle-hole
excitations are accommodated within the independent-particle approximation.
{}From Eq. (\ref{Slater}) it follows that
$n_a(t) = \sum_{k<k_F} |b_k(t)|^2$, and $n_k(t) = \sum_{q < k_F}
|e_{kq}(t)|^2$.

In Fig. \ref{fig6} we plot the time evolution of the entropy in the
independent-particle approximation and in the approximate solution to many-body
model for the case of spinless fermions ($N = 1$) with and without the
new sectors $|a; L,k,q \rangle$ of the Hilbert space, Eq. (\ref{O.PSI}).
We also eliminate excited atomic states in the
many-body equations of motion Eq. (\ref{EOM})
to permit direct comparison of the approximate many-body
solution with the independent-particle solution.
The initial state of the Lithium atom, which is held at fixed position
$z = z_0 = 1.0 \AA$, is a positive ion and the entropy is zero.  As
time advances, this pure state evolves into a mixed one and the entropy
grows.  Several features shown in Fig. \ref{fig6} are
generic for all of the initial conditions and parameters we tested.
First, the entropy increase in the independent-particle case is
comparable to that in the many-body case, even though
the Hilbert space of the independent-particle solution is unrestricted.
This suggests that sectors containing two and more particle-hole pairs,
the ones not present in the variational wavefunction, do not become
significantly populated and can be safely neglected.  Indeed
the number of electron-hole
pairs produced during a collision, estimated in the independent-particle
solution by counting the expected number of particles due to pairs,
$\sum_{k>k_F} n_k$, is typically less than one.
Evidently an infrared catastrophe is avoided: the number of very low
energy excitations is severely limited.
We also see that entropy does not grow monotonically when the
$|a; L, k, q \rangle$ sectors are dropped as the Hilbert space is now too
restricted for phase decoherence to be complete.
Finally, Fig. \ref{fig7} shows how the entropy grows
monotonically in the
many-body solution for the physical case of spinning electrons, $N = 2$, with
both excited neutral and negative ion states now included.

Entropy growth provides clues as to how probability
flows between different sectors within the truncated Hilbert space.
We again turn off the couplings to the negative ion and excited state sectors.
The coarse-grained entropy grows in time as long as
the Hamiltonian is time-independent.  The main direction of the
probability flow corresponds to flow into sectors with larger and
larger phase space.  The phase space corresponding to
$|0 \rangle$ is a single state and is therefore smaller than
the phase space of the $|a; k \rangle$ sector which contains $M$ states.
The $|L,k \rangle$
sector occupies an even larger portion of the phase space consisting of
$M^2$ states.  Finally, sector $|a; L,k,q \rangle$ occupies the
largest portion of the Hilbert space as it comprises $M^3$ states.
Thus, probability which flows in the direction
$|0 \rangle \rightarrow |a;k \rangle$,
$|a; k \rangle \rightarrow |L,k \rangle$, and
$|L,k \rangle \rightarrow |a;L,k,q \rangle$ as shown in Fig. \ref{fig8}(a)
leads to entropy increase while a
reversal of flow would, in general,
lead to a decrease of entropy and is improbable.
As an illustration of probability flow, consider the time dependence of
the occupancies shown in Fig. \ref{fig9}(a).
The initial positive ion $|0 \rangle$ state first dwindles into
a group of neutral $|a; k \rangle$ states because in this case the
surface work-function has been set to the low value of $W = 1.59$ eV.
Later on, the higher order $|L,k\rangle$ and $|a;L,k,q\rangle$ sectors
become partially populated.
Had the initial state been a neutral atom and had the surface work-function
been high,
probability would instead have flowed from the single $|a; k=0 \rangle$ state
diagonally into the ionized states with a particle-hole pair, $|L,k \rangle$.
The alternative ionization route $|a; k = 0\rangle \rightarrow |0 \rangle$
does not increase the entropy
and is negligible compared to the $|a; k \rangle \rightarrow |L,k \rangle$
route.  The approximate solution confirms this scenario in this case as
the particle-hole sector $|L, k \rangle$ dominates the final $A^+$ occupancy.

An important feature of the probability cascade is the
increasing time scale at
which higher-order sectors become populated as seen for instance in
Fig. \ref{fig9}(b).  The atomic occupancies essentially reach their final
values at $t = 7.5 \times 10^{-16}$ sec, despite the fact that probability
continues to flow from $|0\rangle$ to $|L,k\rangle$ and from $|a;k\rangle$
to $|a;L,k,q\rangle$ even at much later times
$t = 20 \times 10^{-16}$ sec.  That the occupancy
of the atomic orbital is unaffected by these subsequent probability flows
to the
higher-order sectors supports the use of the $1/N$ expansion, as the neglect of
terms of order $1/N^2$ and higher should not significantly disturb
observables accessible to experiment.

We next switch back on the coupling to $A^-$ as depicted in Fig. \ref{fig8}(b).
It is now clear why loss-of-memory breaks down
for the $A^-$ initial condition at high work-functions, as shown for instance
in Fig. \ref{fig5}(c).  For high work-functions the final charge state should
be
mostly $A^+$. However, as shown in Fig. \ref{fig8}(b), there is no path of
strictly growing entropy which leads from $A^-$ into any of the
sectors corresponding to $A^+$.  The probability can only flow into
the $|a; L,k,q\rangle$ sector corresponding to a neutral atom and a
particle-hole excitation, and stay there as in Fig. \ref{fig5}(c).
We conjecture that loss-of-memory for this initial condition
can be restored with the inclusion of a new sector corresponding to a
positive ion with two particle-hole pairs, which appears
at second order in $1/N$ as shown in Fig. \ref{fig8}(b).
The probability may
then cascade diagonally down from $A^-$ to $A^+$ with increasing entropy at
each step. The reason this sector has not been included is pragmatic:
it is labeled by four momenta indices, and the computational power
required to solve
$O(M^4)$ differential equations versus $O(M^3)$ at the current level
of approximation would be excessive.

Up until now we have focused on the static problem of an atom at fixed
distance from the surface.  We now return to the dynamical problem.
Consider the positive ion initial condition (1)
and a surface of intermediate work-function.  Away from the surface,
the atomic level lies below the Fermi energy and the atom neutralizes via
the $|0 \rangle \rightarrow |a; k \rangle$ path as shown in Fig. \ref{fig8}(a).
Close to the surface the level is image shifted above the Fermi energy
and probability flows back into the positive ion sector $|L,k \rangle$.
On the outgoing leg of the trajectory, the atomic level shifts back below
the Fermi energy and the atom again neutralizes by filling up
$|a; L,k,q \rangle$ sector.  As the image shift is a monotonic function
of distance $z$, higher order sectors do not become populated significantly
during the atom-surface collision, as this would require more than two level
crossings.  In the dynamical problem there is back flow of probability,
manifested as a decrease in entropy along part of the outgoing trajectory as
shown in Fig. \ref{fig10}.  The decrease in entropy does not contradict
the quantum generalization of the H-theorem as the Hamiltonian now
depends explicitly on time.  Even though the incoming Li$^+$ ion is completely
neutralized during its encounter with the $W = 1.59$ eV surface, the
probability for exciting an electron into one of the unoccupied levels above
the Fermi energy is only 0.098 in the independent-particle approximation.
Thus the probability for the creation of
a particle-hole excitation is comparably small.

Two conclusions should be emphasized.  First, the introduction of a
coarse-grained entropy permits a quantitative
understanding of the loss-of-memory process and elucidates the origin of
irreversibility in charge-transfer.   It also facilitates analysis of the 1/N
expansion.  Second, truncation of the Hilbert space at first order in $1/N$ in
most instances suffices for the dynamical charge transfer problem
as the probability flow does not significantly populate higher order sectors
during the course of the atom-surface interaction.
This conclusion is supported by the
independent-particle approximation which shows that less than one particle-hole
pair is produced under typical conditions.

\section{Auger Processes}
\label{sec:auger}

We can take advantage of the newly added extension to the
Hilbert space to include Auger charge transfer processes in addition to
the resonant processes considered up until now.  It has
been a long standing question\cite{Amos} whether or not
Auger processes are of comparable importance to resonant charge transfer.
We show here that at least at low surface work-functions,
Auger transitions may be required to obtain an accurate description of
experiments involving Lithium bombardment of Copper surfaces with alkali
overlayers.  The measured yield of excited neutral Li(2p) atoms
grows at the very lowest work-functions when the incoming kinetic energy of the
Li$^+$ ion is $100$ eV.  At kinetic energies of $400$ eV, however,
this feature appears to be absent.\cite{Behringer}

In a typical Auger process
an electron from one of the filled states below the Fermi energy
of momentum $q$ transfers non-resonantly
into the atomic orbital $a$,
while a second metal electron below $\epsilon_{F}$
of momentum $k$ is promoted to a state of momentum $L$ of higher energy.
Within the truncated Hilbert space transitions with
$L >~ k_{F}$ and $k, q <~ k_F$ couple the $|0\rangle$ sector
to the $|a; L,k,q\rangle$ sector as shown in Fig. \ref{fig11}(a).
Two other possible Auger processes
are shown schematically in Fig. \ref{fig11}(b) and (c).
In these cases one of the metal electrons hops onto
the atomic orbital while the other remains below
the Fermi level but fills up a hole
which was already present.  These transitions couple the neutral
$|a; k \rangle$ sector to the negative $|k,q\rangle$ sector.
Other Auger processes which we do not consider here include Auger de-excitation
of the neutral atom and transitions between the $|L,q\rangle$ and
$|a; k\rangle$ sectors.

It is straightforward to include new terms in the Hamiltonian Eq. (\ref{O.H})
which correspond to these processes:
\begin{equation}
H^{Aug}(t) = {{1}\over{N}}~ \sum_{L,k,q} V^{A(1)}_{a;Lkq}(z)~ \hat{P}_1~
c_L^{\dag \alpha} c_{k \alpha} c_a^{\dag \beta} c_{q \beta}
+ {{1}\over{\sqrt{N}}}~ \sum_{l,k,q} V^{A(2)}_{0;lkq}(z)~ \hat{P}_2~
c_l^{\dag \alpha} c_{k \alpha} c_0^{\dag \beta} c_{q \beta} + H.c.
\label{AugerHam}
\end{equation}
We use the same notation here as in Eq. (\ref{O.H}) except that now the
sum over momenta indices is restricted to states either above or below
the Fermi energy, depending on whether the index is a capital or
lower case letter.
We have normalized the couplings $V^{A(1,2)}$ differently to account
for the $N$ species of spins.  For Auger transitions $V^{A(2)}$
from the neutral $|a; l\rangle$ sector to the negative $|k,q\rangle$ sector,
a pre-existing hole in the metal of momentum $l$ and specific spin must be
filled.  This is not the case for Auger transitions $V^{(1)}$
from a positive ion to
a neutral atom which involve the creation of two new holes of any spin; hence,
the matrix element must be reduced by an additional factor of $1/\sqrt{N}$
to make the $N \rightarrow \infty$ limit well defined.
As we work in the restricted Hilbert space
defined previously, projection onto singly and doubly occupied atomic sectors
occurs automatically
and we may drop the projection operators $\hat{P}_{1,2}$ in the following
equations.  Before we proceed further it is useful to separate the Auger
Hamiltonian into
symmetric and antisymmetric parts (with respect to interchange of
the momenta indices $k$ and $q$) to accord with Eq. (\ref{N.Sector}).
We also ignore the momentum dependence of the Auger
matrix elements by setting $V^{A(1,2)}_{a;lkq}(z) = V^{A(1,2)}_{a}(z)$.
This approximation, like the one already imposed on the resonant
matrix elements,
can easily be relaxed to incorporate more complicated momentum dependence.
With this assumption, the antisymmetric part of $H^{Aug}(t)$ vanishes, and
\begin{eqnarray}
H^{Aug}(t) &=& {{1}\over{N}}~ \sum_{L, k > q}
V^{A(1)}_{a}(z) (c_L^{\dag \alpha} c_{k \alpha} c_a^{\dag \beta} c_{q \beta}
+ c_L^{\dag \alpha} c_{q \alpha} c_a^{\dag \beta} c_{k \beta})
\nonumber\\
&+& {{1}\over{\sqrt{N}}}~ \sum_{l, k > q}
V^{A(2)}_{0}(z) (c_l^{\dag \alpha} c_{k \alpha} c_0^{\dag \beta} c_{q \beta}
+c_l^{\dag \alpha} c_{q \alpha} c_0^{\dag \beta} c_{k \beta})
+ H.c.
\label{AugerS}
\end{eqnarray}
Adding the Auger Hamiltonian to the resonant one, Eq. (\ref{O.H}),
and projecting the resulting
Schr{\"o}dinger equation onto each sector of the Hilbert space, we obtain
the following terms to be added to
the equations of motion Eq. (\ref{EOM}):
\begin{eqnarray}
i{d\over{dt}}F &=&      ...  \  +  \sqrt{2 (1 - 1/N)}~
\sum_{a;\ L,k>q} V^{A(1)*}_{a}(z)~
{\exp}\{-i[(\epsilon_L  - \epsilon_k  - \epsilon_q )t  + \phi_a(t)]\}\
S_{a; Lkq}.
\nonumber\\
i{d\over{dt}}B_{a;l} &=&   ... \ -2 \delta_{a,0}~ \sqrt{1 - 1/N}~
\sum_{k>q} V^{A(2)*}_{0}(z)~
{\exp}\{i[(\epsilon_k  + \epsilon_q  - \epsilon_l - 2\epsilon^{(2)}_0 - U) t
+ \phi_0(t)]\}\ D_{kq}.
\nonumber\\
i{d\over{dt}}E_{Lk} &=&   ...
\nonumber\\
i{d\over{dt}}D_{kq} &=&   ... \ -2 \sqrt{1 - 1/N}~
\sum_{l}V^{A(2)}_{0}(z)~
{\exp}\{-i[(\epsilon_k  + \epsilon_q  - \epsilon_l - 2\epsilon^{(2)}_0 - U) t
+ \phi_0(t)]\}\ B_{0;l}.
\nonumber\\
i{d\over{dt}}S_{a; Lkq} &=&  ... \  + \sqrt{2 (1 - 1/N)}~ V^{A(1)}_{a}(z)~
{\exp}\{i[(\epsilon_L  - \epsilon_k  - \epsilon_q )t + \phi_a(t)]\}\ F.
\nonumber\\
i{d\over{dt}}A_{a; Lkq} &=&  ...
\label{EOMAg2}
\end{eqnarray}
Here the ellipses denote all of the terms in the original equations of
motion, Eq. (\ref{EOM}), which remain unchanged.  Like Eq. (\ref{EOM}),
the new equations of motion, Eq. (\ref{EOMAg2}), are exact in the
$N \rightarrow \infty$ limit.  Higher order terms are suppressed by powers
of $1/N$.

Before we proceed with the solution of the above system of equations,
we must find reasonable values for the Auger matrix $V^{A(1,2)}_{a}(z)$.
Adopting the same parametrization scheme for $V^{A(1,2)}_{a}(z)$
as in the case of the resonant couplings, we assume that the couplings fall
off exponentially fast at large distances from the surface and saturate close
to it.  As now there are four overlapping wave functions in the matrix element
(compared to two in the resonant case) we expect the coupling to fall off
roughly twice as fast away from the surface\cite{Fonden}.
We obtain the couplings from the corresponding atomic half-widths
by using the Fermi Golden Rule.  For the neutral atom:
\begin{equation}
{\Delta_a}^{A(1)}(z) = \pi \sum_{k,q,L} |V^{A(1)}_a(z)|^2~
\rho~ \delta_{\epsilon_L + \epsilon_a^{(1)} - \epsilon_k - \epsilon_q, 0}\ ,
\label{FGRgen1}
\end{equation}
and for transitions to the negative ion state:
\begin{equation}
{\Delta_0}^{A(2)}(z) = \pi \sum_{k,q,l} |V^{A(2)}_0(z)|^2~
\rho~ \delta_{\epsilon_0^{(1)} - \epsilon_l,
2\epsilon_0^{(2)} + U - \epsilon_k - \epsilon_q}
\label{FGRgen2}
\end{equation}
where again the density of states
for a conduction band described by a set of $M$ equidistant levels spaced
$D/M$ apart is given by $\rho = M/D$. From Eq. (\ref{FGRgen1}), it follows
that:
\begin{equation}
{\Delta_a}^{A(1)}(z) = {1\over 2}\pi (M/D)^3 {|V^{A(1)}_{a}(z)|}^{2}
{(\epsilon_a^{(1)} - \epsilon_F)}^2~ \theta(\epsilon_F - \epsilon_a^{(1)})\ .
\label{FGRAg}
\end{equation}
In Eqs. (\ref{FGRAg}) the Auger rate is proportional to
$(\epsilon_a - \epsilon_F)^2$.  The rate, like the inverse lifetime of
a quasiparticle in a Landau Fermi liquid,\cite{Negele}
drops rapidly as the phase space available to particle-hole pair excitations
decreases.  For $|\epsilon_a^{(1)} - \epsilon_F| = D$, however,
\begin{equation}
{\Delta_a}^{A(1)}(z) = {\pi \over 2} {{M^3}\over D} {|V^{A(1)}_{a}(z)|}^{2}\ .
\label{FGRAgref}
\end{equation}
Assuming that the holes are confined to the energies just below the Fermi
level,
and assuming that
the energy difference between the negative ion state and the neutral
ground state equals the half-bandwidth $D$, we also find:
\begin{equation}
{\Delta_0}^{A(2)}(z) = {\pi \over 2} {{M^3}\over{D}} {|V^{A(2)}_{0}(z)|}^{2}\ .
\label{FGRAg-}
\end{equation}

To be concrete, we choose as trial
parameters for ${\Delta_a}^{A(1,2)}$ those listed in
Table \ref{table2} and obtain the matrix elements $V^{A(1,2)}$ from
Eqs. (\ref{FGRAgref}) and (\ref{FGRAg-}).  Of course, the matrix elements
themselves, not the widths, are of fundamental importance in the many-body
theory.  For instance, the sign of the couplings is important; we choose
$V^{A(1)} > 0$ and $V^{A(2)} < 0$ so that the Auger processes interfere
constructively with the resonant ones.
The Fermi Golden rule then determines the
magnitude of the matrix elements in an approximately correct, but
M-independent, way.  The results of the dynamical calculation for Lithium
which includes both resonant and Auger charge transfer are presented in
Fig. \ref{fig12}.  The Auger couplings have been chosen to be
sufficiently small so
that the peak in the excited neutral Li(2p) occupancy remains at work-function
$W \approx 2.8$ eV.  Now, however, there is a second upturn at
$W \approx 1.5$ eV in qualitative agreement with experiment\cite{Behringer}.
These features are robust as the two upturns remain even when the Auger
rates are doubled or halved.
To understand the origin of the second upturn at low work-functions,
we first review the explanation for the existence of a peak at $W = 2.8$ eV
when there is only resonant charge transfer\cite{Behringer}.
Dynamical competition between Li(2p) and Li(2s) states is the key to
understanding the photon peak.  As the atom bounces away from the surface
the coupling between the surface and the atom falls off faster
for the Li(2s) state than for the Li(2p) state because the
Li(2p) orbital is larger and of higher energy than the Li(2s) orbital.
At the highest work
function values, the energy of Li(2p) state lies above the Fermi level
at all distances from the surface and is unoccupied.  However, as the
work-function is lowered, Li(2p) state begins to cross the Fermi level
at closer distances where its coupling to the surface is appreciable while
the coupling to Li(2s) is still small.  For this intermediate range of
work-functions, the Li(2p) state becomes populated on the outgoing
leg of the trajectory despite the fact that it is always
energetically less favorable than Li(2s). As the work-function drops
further, the Fermi level crossing for the Li(2p) state occurs at distances
for which coupling to Li(2s) is appreciable.  Now the Li(2p)
state yields its occupancy to the lower energy Li(2s) state.

Consider what happens when the Auger coupling is turned on.
At the very lowest work-functions, Auger transitions between the Li(2p) state
and the metal occur more frequently because the phase space for these processes
grows rapidly as $(\epsilon_a^{(1)} - \epsilon_F)^2$ increases.
As we expect the $\Delta_{2s}(z)$ Auger rate to fall off more rapidly
away from the surface than the $\Delta_{2p}(z)$ Auger rate,
the picture outlined above in the resonant case
simply repeats itself and there is a second
upturn in Li(2p) occupancy. However, as evident from Fig. \ref{fig12},
the second rise in Li(2p) occupancy is not noticeable at the higher incoming
kinetic energy of $400$ eV.  In the $100$ eV case the atom
moves only half as fast as in $400$ eV case and there is sufficient
time for an electron to make an Auger transition
to the excited neutral Li(2p) state.  At higher velocities there is not
enough time for an Auger transition to occur.

While the model developed here reproduces the upturn in the
excited neutral Li(2p) occupancy at the very lowest work-functions, it is
only one of several possible explanations for the feature.  Two
difficulties impede further progress.  Experimentally, it is hard to measure
absolute yields of ejected Auger electrons.  Relative yields, as measured
in Auger spectroscopy\cite{Brenten}, provide little guidance.
Most of the
Auger electrons are promoted to unoccupied metal states instead of ejected
from the surface.  Indirect probes, such as the formation of excited states,
appear to be the only way to gauge the relative importance of Auger processes.
Theoretically, Auger matrix elements cannot be computed
accurately because Auger transitions are driven by many-body
correlations\cite{Monreal}.  The parameters for the Auger rates
listed in Table \ref{table2} are at best just an educated guess.
Indeed, calculations to date have focused on
Auger widths as opposed to matrix elements which are of more fundamental
importance.  Either constructive or destructive interference
with resonant processes is possible, but only a full microscopic calculation
of both types of matrix elements can determine the relative sign.

\section{Conclusions}
\label{sec:conc}
In this paper we described a generalized Newns-Anderson model
of charge transfer and its systematic solution based on a $1/N$ expansion.
We went beyond earlier work by including new sectors in the Hilbert
space and showed that loss-of-memory was improved.
We analyzed the effect of the truncation of the Hilbert space
on loss-of-memory by studying entropy production both within the approximate
solution to the many-body theory and also within the independent-particle
approximation.
This analysis showed how the $1/N$ truncation scheme works in dynamical
problems.  In most cases, higher order sectors can
be neglected as less than one particle-hole pair is
produced during the atom-surface collision.  This conclusion was supported
by the independent-particle approximation.  Despite the fact that an unlimited
number of particle-hole pairs can be accommodated within this approximation,
typically less than one is created during an atom-surface interaction.
The production of entropy during the collision demonstrates the
irreversibility of the interaction: at velocities of experimental interest,
information about the initial state of the incoming atom is dissipated.

We included Auger processes and showed that an experimentally observed
upturn in photon yield
due to the formation of excited Li(2p) atoms at the very lowest
work-functions can be
explained in terms of competition between the Li(2s) and Li(2p) states and
the rapid growth at low work-functions in the phase space for Auger
transitions.
Finally, we examined whether Kondo effects are experimentally accessible,
and concluded that it will be extremely difficult to separate the small
predicted effects from other, more mundane, nonlinearities.

\section{Acknowledgments}

We thank Ernie Behringer, Dennis Clougherty, Barbara Cooper,
Dan Cox, David Goodstein, Peter Nordlander, Hongxiao Shao,
Chris Weare, and especially Eric Dahl for fruitful discussions.
This work was supported in part by the National Science Foundation through
Grants Nos. DMR-9313856 and DMR-9357613 and by a grant from the Alfred
P. Sloan Foundation.


\newpage

\begin{table}
\caption{Parameters appearing in Eq. 6 which characterize the resonant
half-widths for
different atomic states of Lithium. All parameters are in atomic units.}
\vskip0.1in

\label{table1}
\begin{tabular}{dddd}
{ \sc atomic state } & { $\Delta_{0}$ } & { $\alpha$ } & {$\Delta_{sat}$} \\
\hline
{$Li^{0}(2s)$}  & 2.23 & 0.86 & 0.04 \\
{$Li^{0}(2p_z)$}  & 0.70 & 0.54 & 0.04 \\
{$Li^{-}(2s^2)$}  & 0.18 & 0.38 & 0.05 \\
\end{tabular}
\end{table}

\begin{table}
\caption{Parameters which characterize the Auger half-widths for Lithium.
All parameters are in atomic units.}
\vskip0.1in

\label{table2}
\begin{tabular}{dddd}
{\sc atomic state } & { $\Delta_{0}$ } & { $\alpha$ } & { $\Delta_{sat}$ }\\
\hline
{$Li^{0}(2s)$}  & 14 & 1.8 & 0.7 \\
{$Li^{0}(2p_z)$}  & 10 & 1.3 & 0.6 \\
{$Li^{-}(2s^2)$}  & 5 & 1.4 & 0.3 \\
\end{tabular}
\end{table}

\newpage

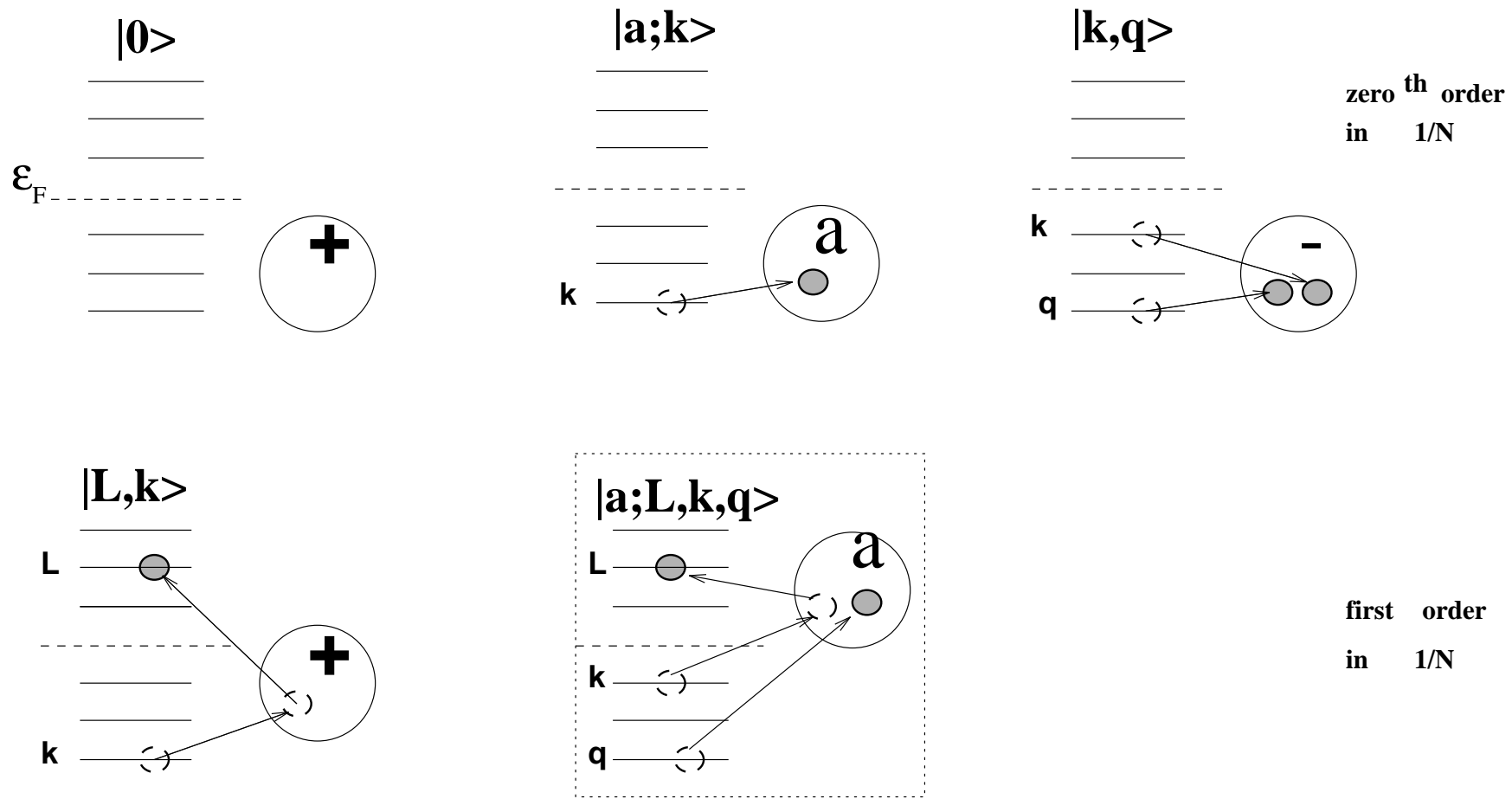
\begin{figure}
\caption{Schematic of the different sectors of the Hilbert space
up to order $1/N$. The new sector is highlighted in the box.  Still missing
at $O(1/N)$ are amplitudes for a negative ion with two holes plus a
particle-hole pair in the metal.}

\label{fig1}
\end{figure}

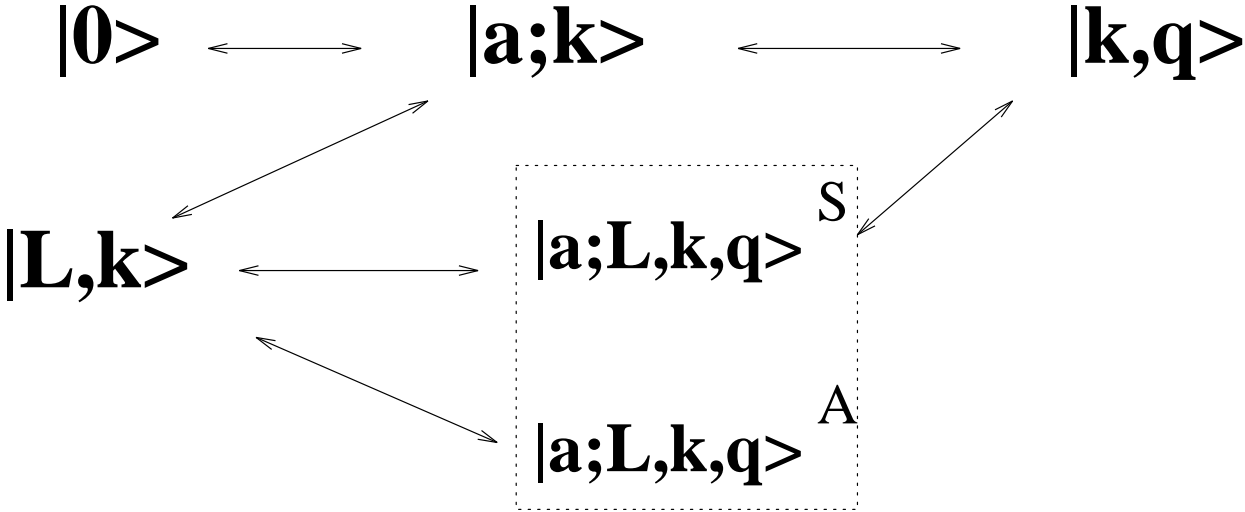
\begin{figure}
\caption{Resonant charge transfer couples the different sectors of the
truncated Hilbert space indicated by the arrows.}
\label{fig2}
\end{figure}

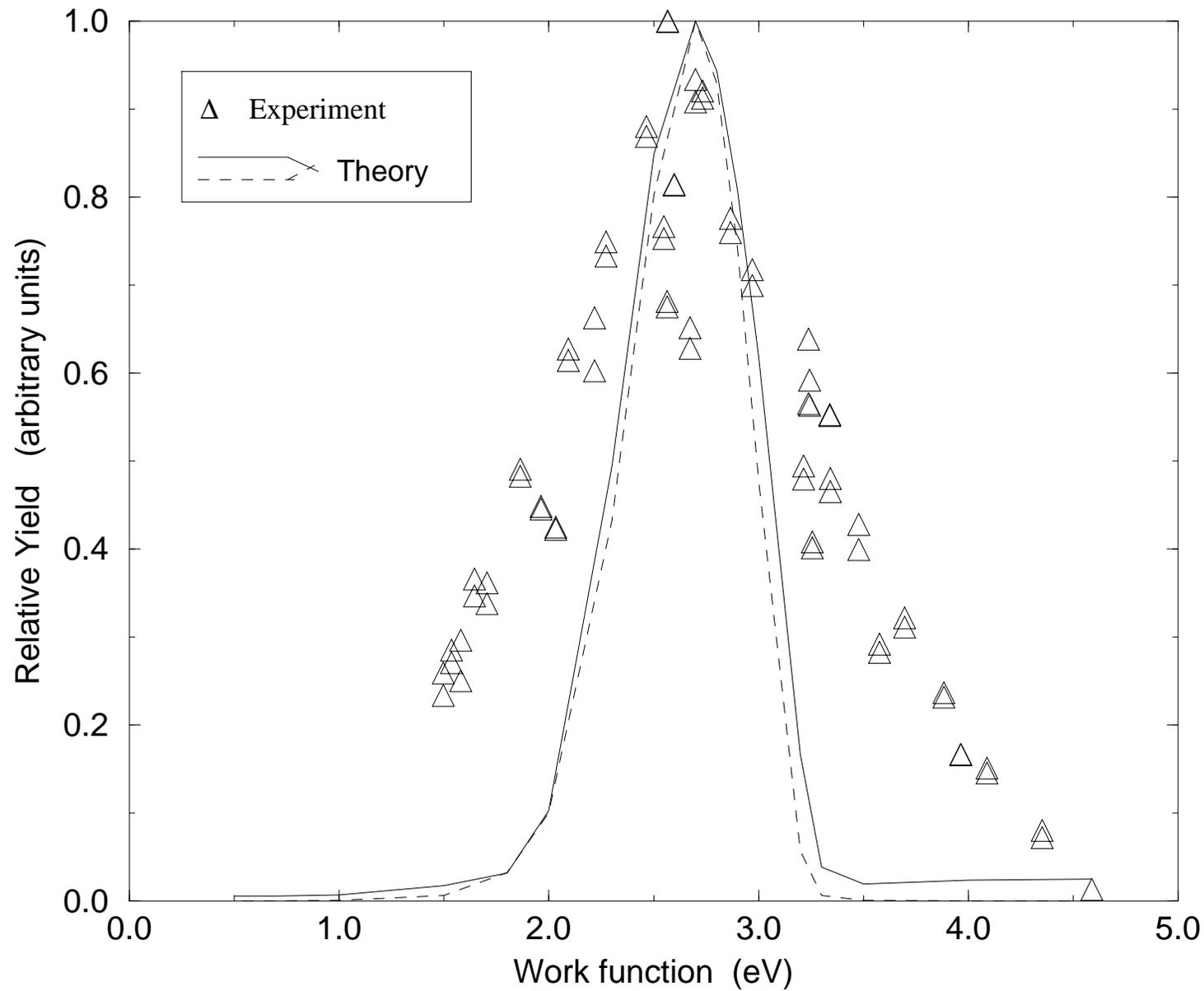
\begin{figure}
\caption{The measured and predicted normalized yields of the
excited neutral atom
Li(2p) (triangles) versus the surface work-function $W$.
In the experiment,
$Li^{+}$ is incident on K/Cu(001) with initial kinetic energy
$E_0 = 400$ eV. The peak occurs at $W \approx 2.8$ eV.
Solid and the dashed lines are the results of the improved
approximate solution of the many-body model ($N=2$) for two different initial
conditions.  In this case the band consists of $M = 30$ states
above and $30$ states below the Fermi surface with a half-bandwidth
of $D = 4$ eV.  The atomic level width parameters are given in Table I.
For the initial condition of the equilibrium ground state (solid line)
the trajectory begins from the point of closest approach ($z_0 = 1.0 \AA$ )
with an outward velocity given by $u_f = 0.03$ a.u.
For the initial condition of a neutral atom (dashed line), the trajectory
starts at $z_f = 20.0 \AA$ with
initial velocity of $u_i = 0.04$ a.u., bounces at $z_0 = 1.0 \AA$,
and leaves the surface with a lower outward velocity of $u_f = 0.03$ a.u.
The experimental and theoretical yields,
which agree in magnitude, are here normalized to unity. We attribute the
broader width of the experimental peak to inhomogeneities in the surface
potential due to the K adsorbates.}
\label{fig3}
\end{figure}

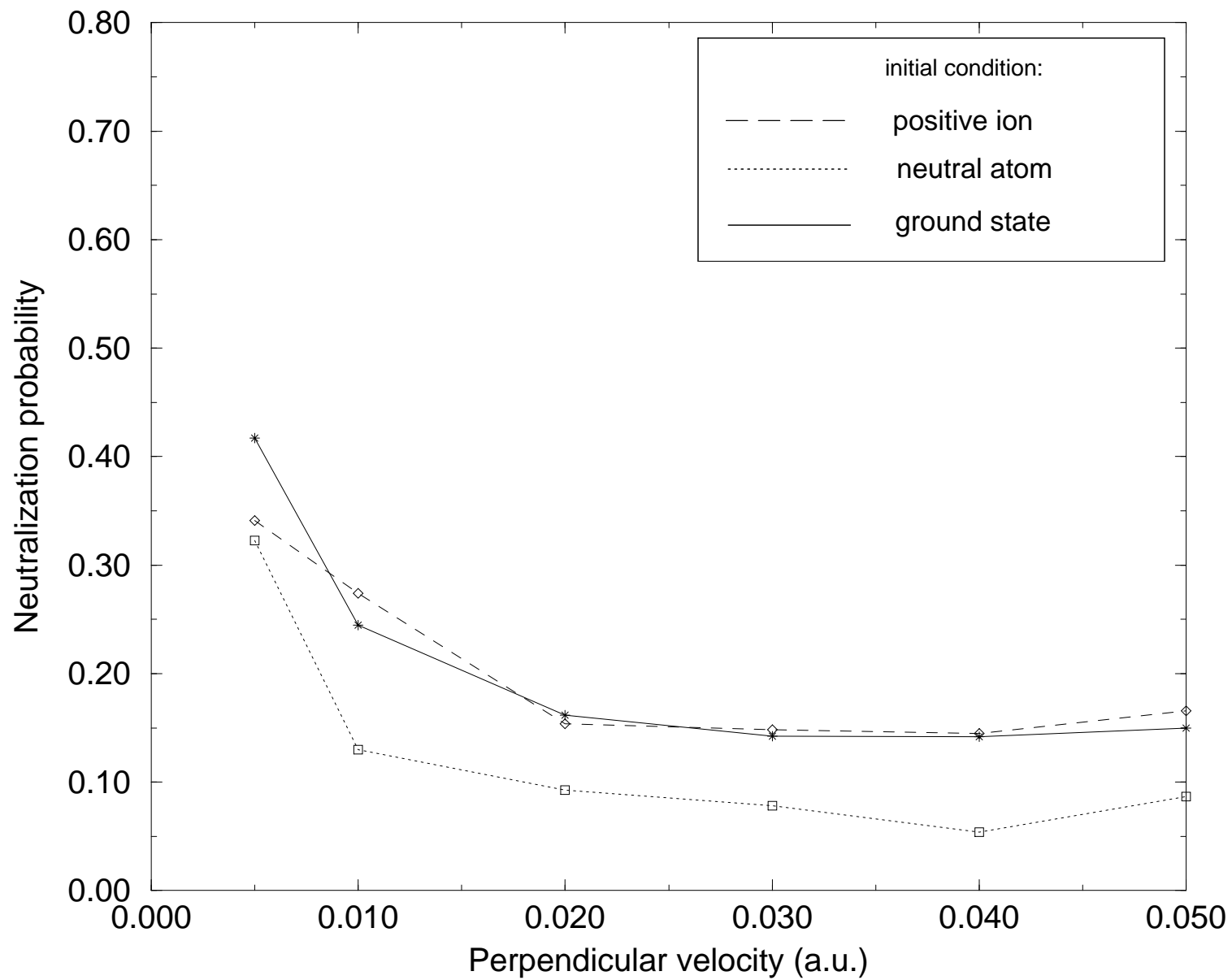
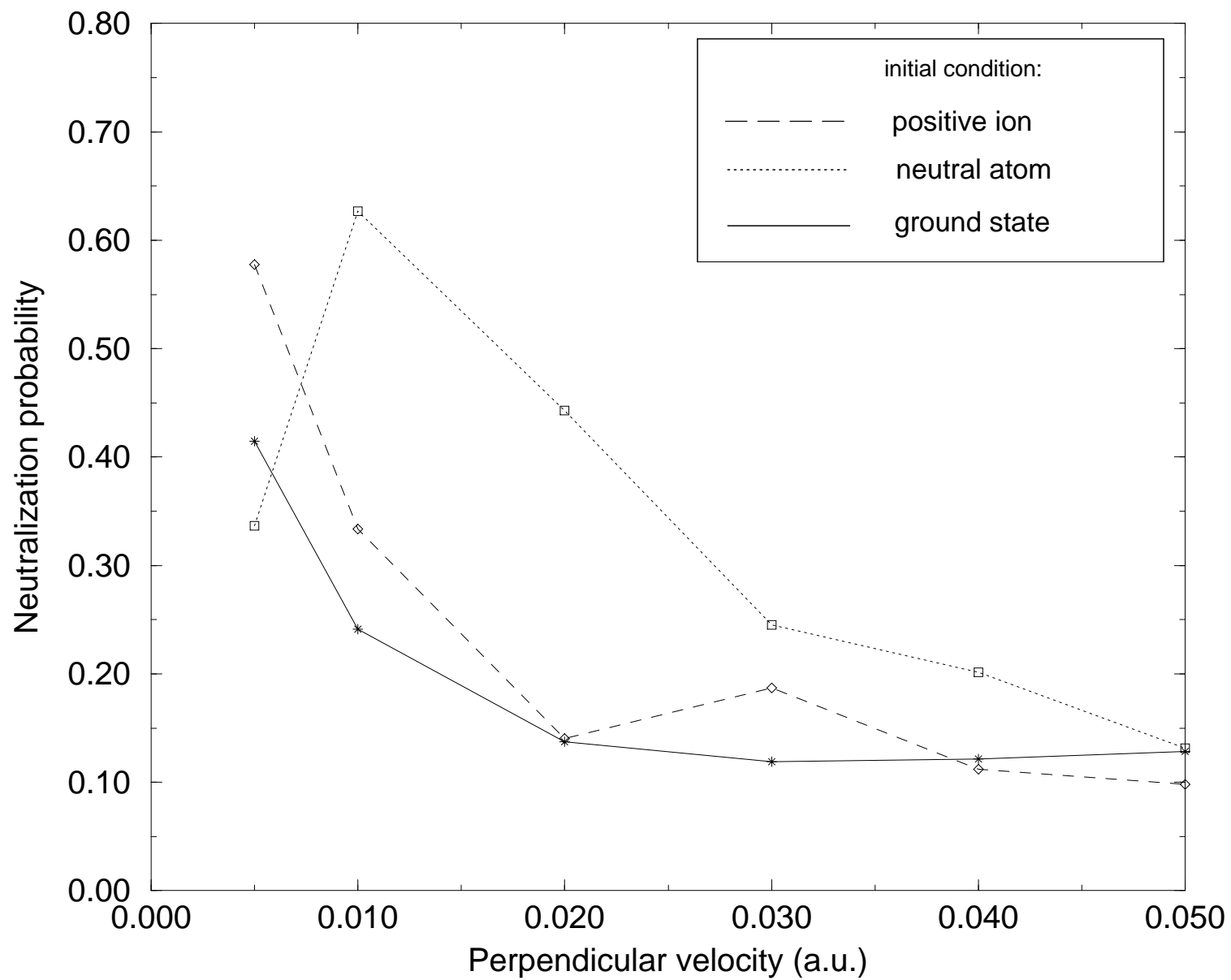
\begin{figure}
\caption{(a) The calculated neutralization probability for Lithium
($N = 2$) as a function
of outgoing velocity $u_f$ using the improved approximate solution.  Three
different initial conditions are examined.  The incoming
velocity, except in the case of the ground state initial condition, is
given by $u_i = (4/3) u_f$.  The surface work-function
is $W = 4.59$ eV, corresponding to a clean Cu(001) surface. (b) Same as (a)
but using the smaller variational Hilbert space and equations of motion of
[I].}
\label{fig4}
\end{figure}

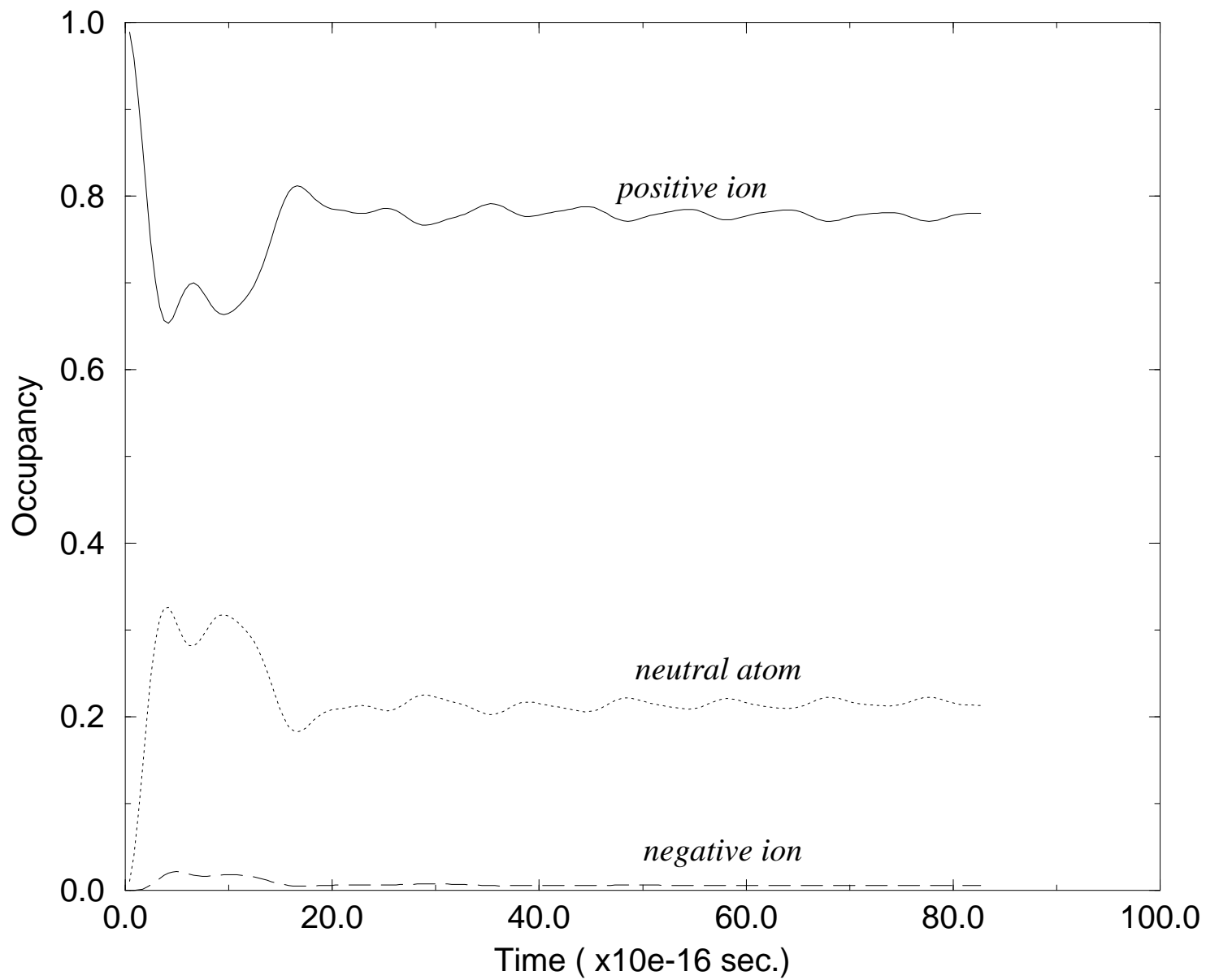
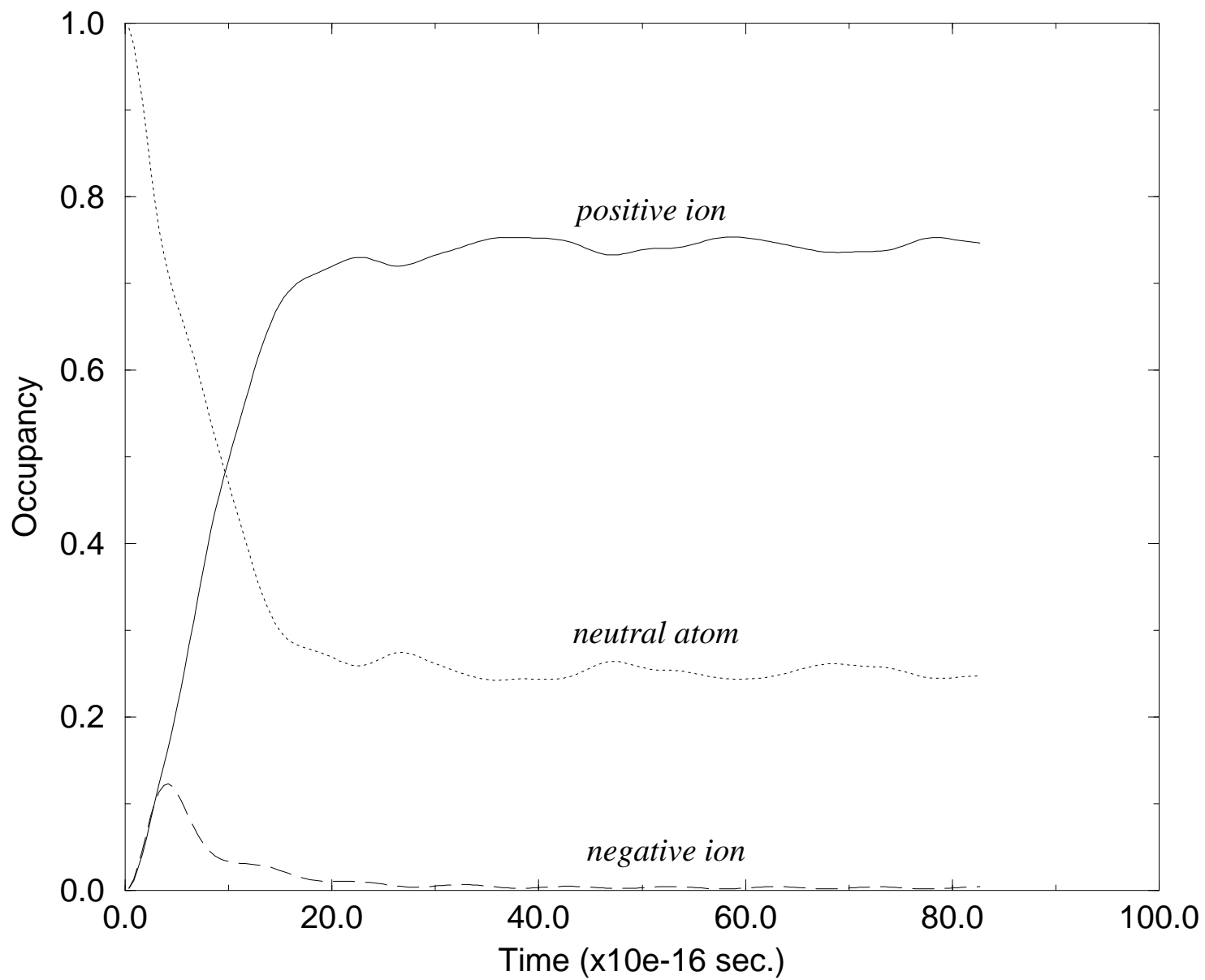
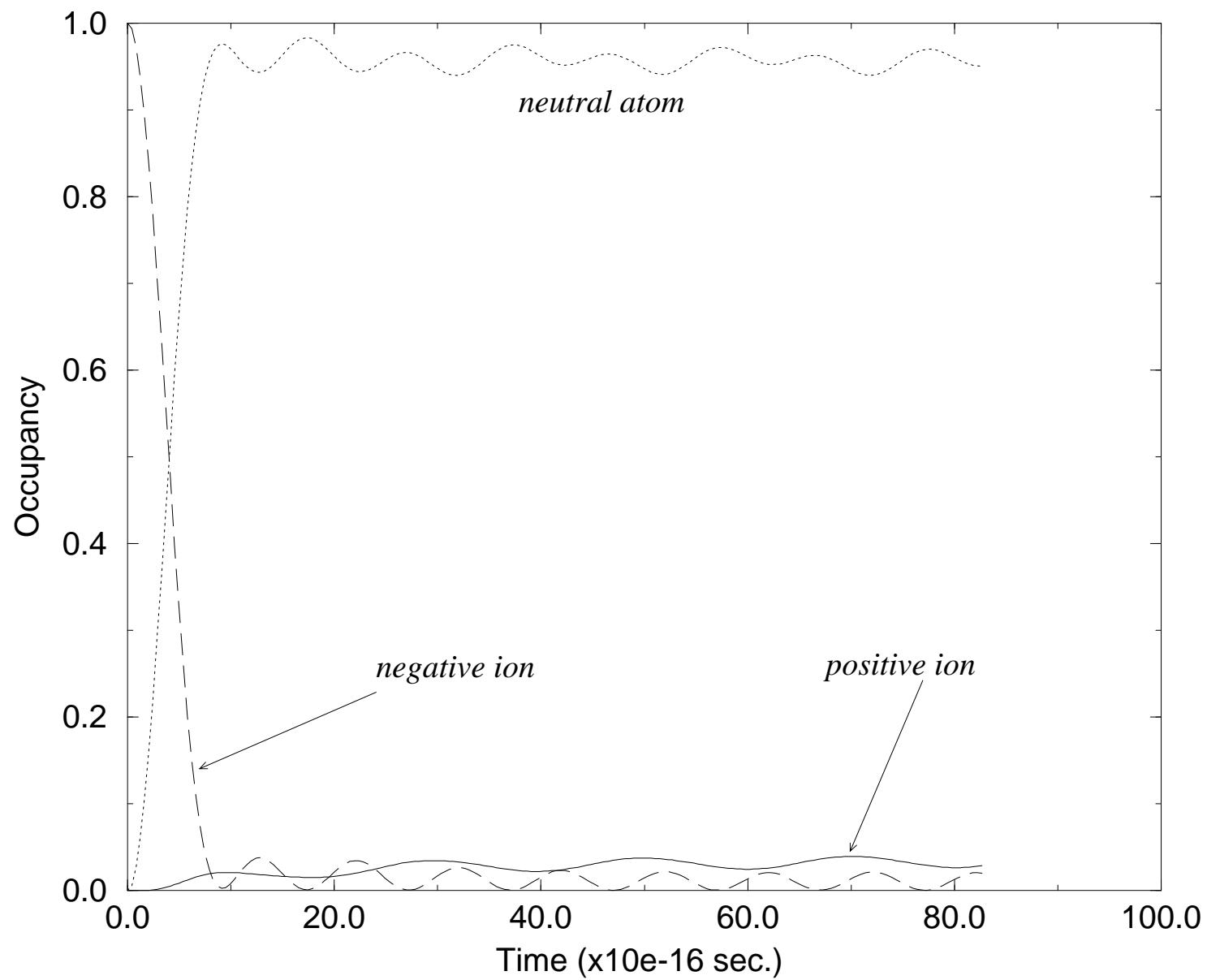
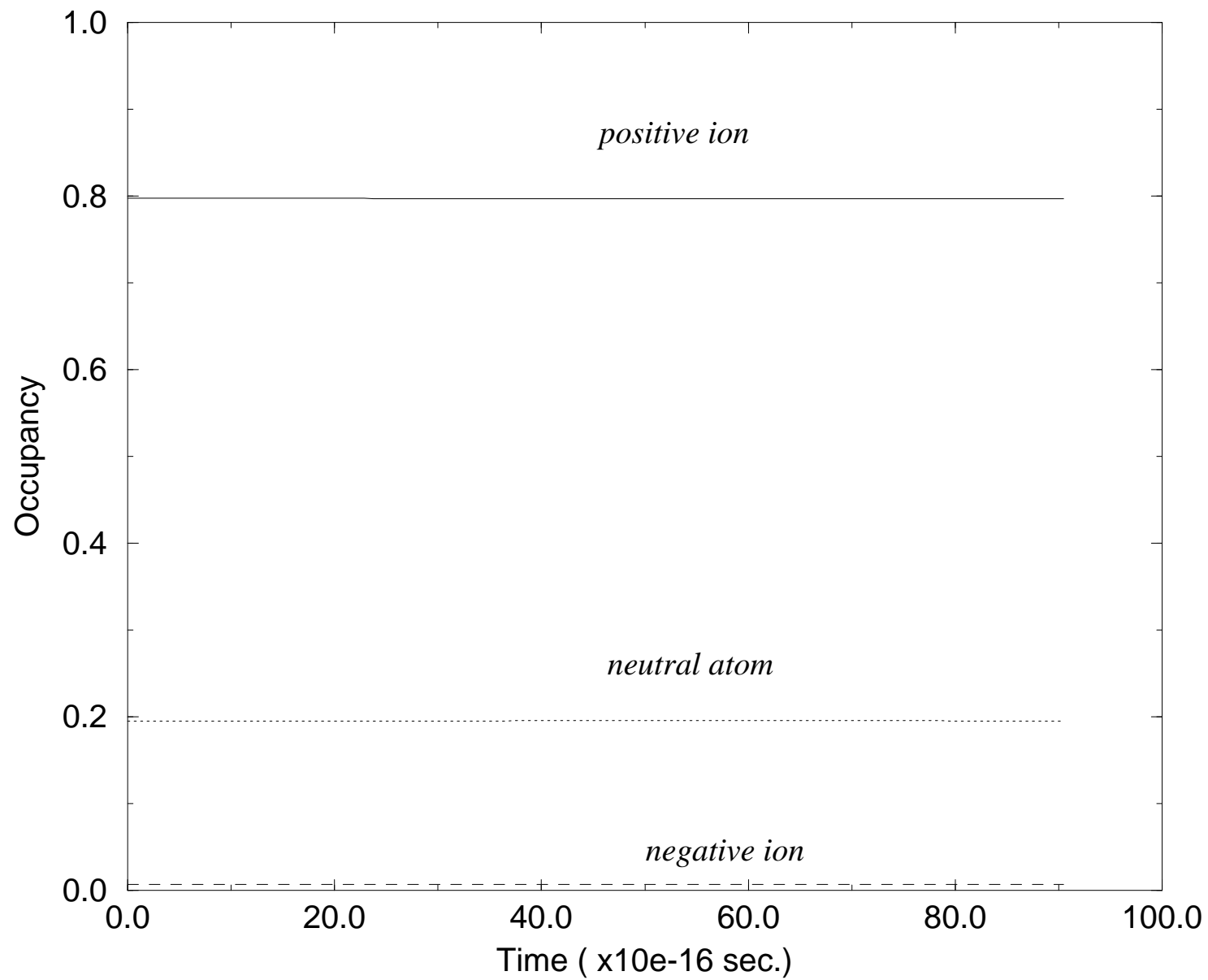
\begin{figure}
\caption{The occupancies of the different charge states $A^{+}$, $A^{0}$
and $A^{-}$ as a function of time for {\it fixed} atomic position $z = z_0$.
A Lithium atom ($N = 2$) interacts with a metal
surface of work-function $W = 4.59$.
Time evolution begins from each of the following four initial conditions:
(a) Positive ion A$^+$ at $z = 1 \AA$.
The final occupancies (at $t = 8.2 \times 10^{-15}$ sec.)
are $P^+ = 0.7806$, $P^0 = 0.2134$ and $P^- = 0.0058$.
(b) Neutral atom A$^0$ at $z = 1 \AA$.
The final occupancies are $P^+ = 0.7471$, $P^0 = 0.2480$
and $P^- = 0.0047$.
(c) Negative ion A$^-$ at $z = 1 \AA$.
The final occupancies are $P^+ = 0.028$, $P^0 = 0.9508$
and $P^- = 0.0204$.
(d) Equilibrium ground state at $z = 1 \AA$.
The final occupancies are $P^+ = 0.7965$, $P^0 = 0.1983$
and $P^- = 0.0071$.}
\label{fig5}
\end{figure}

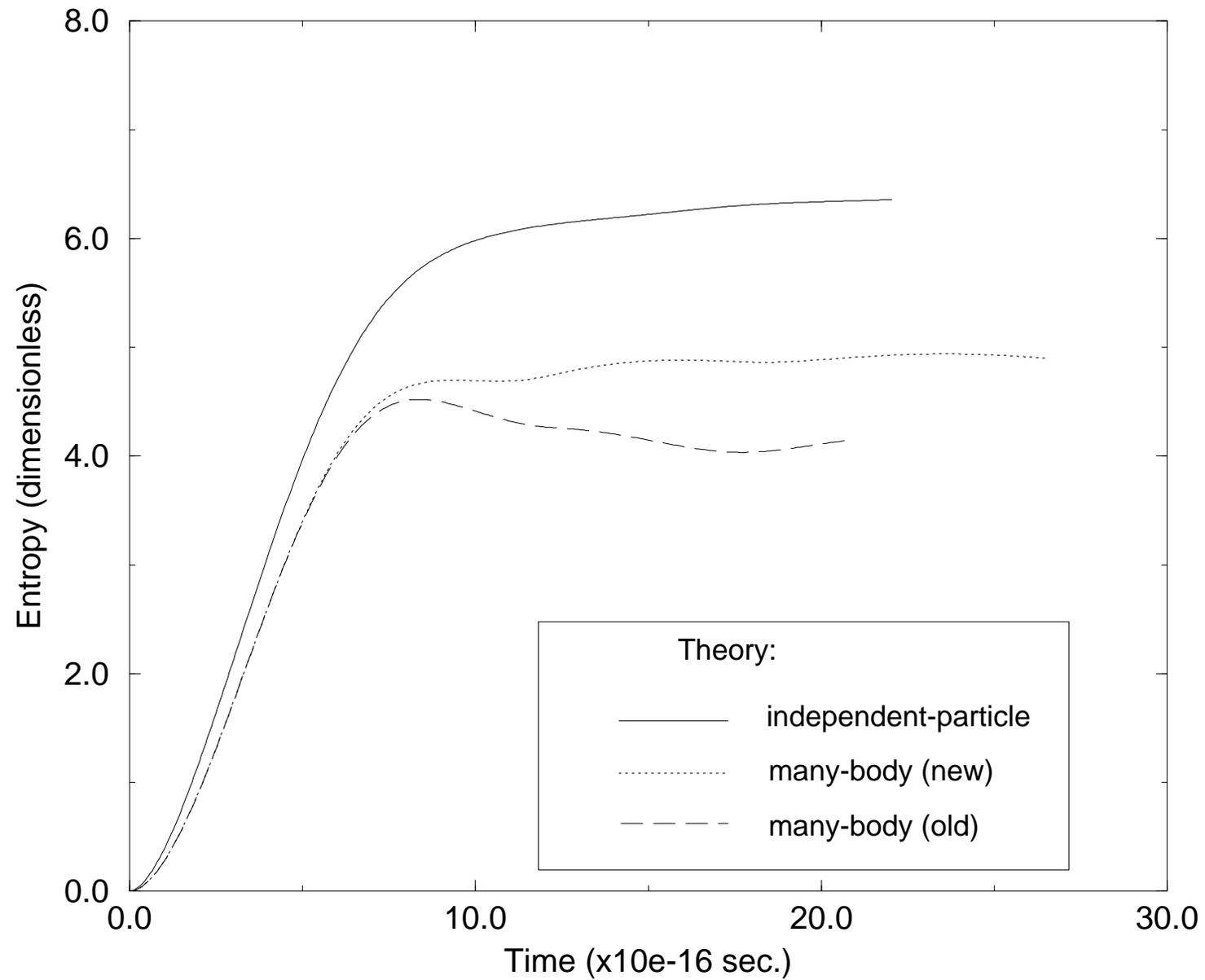
\begin{figure}
\caption{Time-evolution of the dimensionless coarse-grained entropy
$S_{cg}(t)$ for a {\it fixed} atomic position $z = 1 \AA $.
Initially, at $t = 0$, the Lithium atom is a positive ion.  It then interacts
with a metal surface of work-function $W = 1.59$.
The independent-particle and the two approximate many-body solutions
are compared for the case of spinless electrons, $N=1$.}
\label{fig6}
\end{figure}

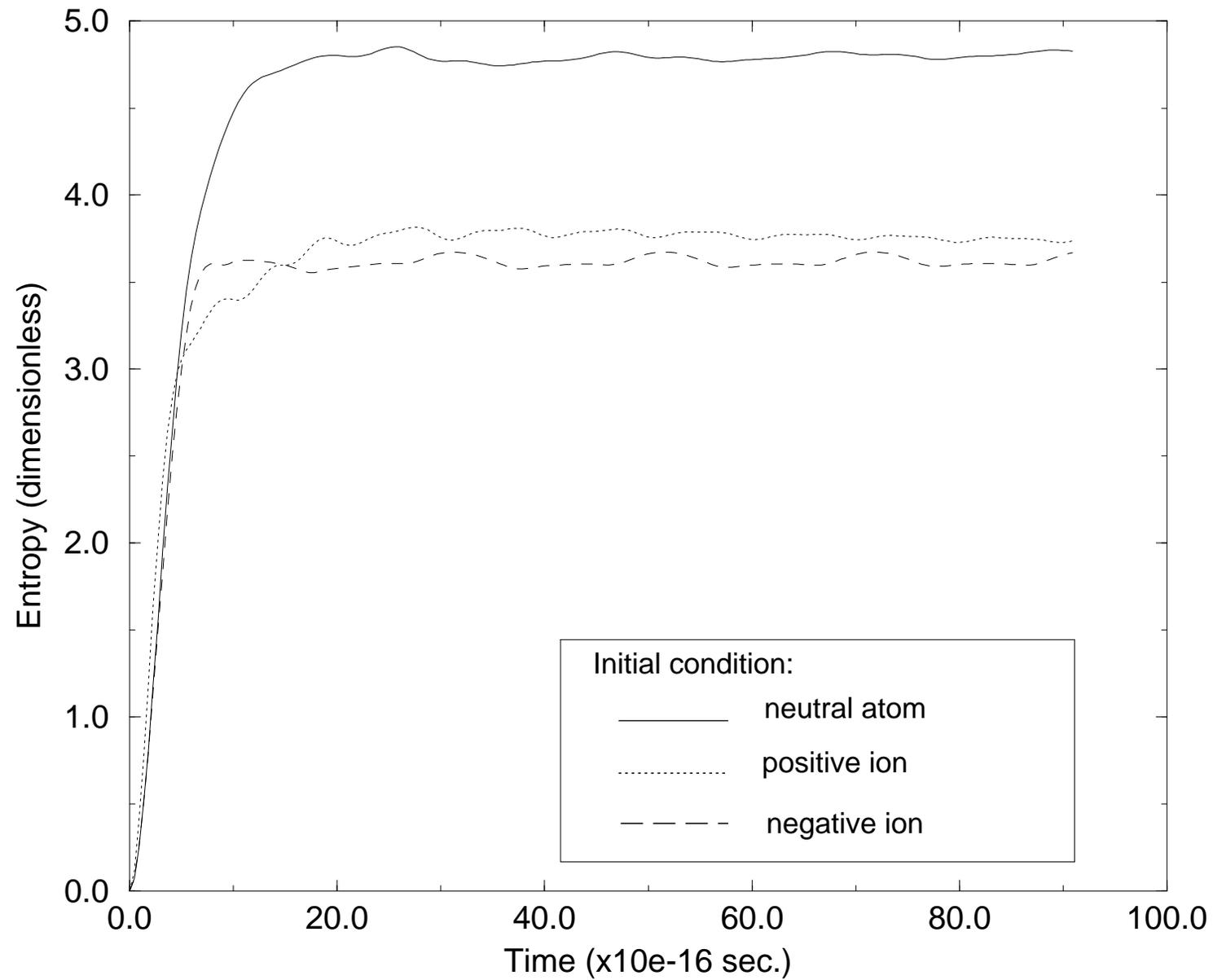
\begin{figure}
\caption{Coarse-grained entropy $ S_{cg}(t)$ as a function of time
for the approximate many-body solution in the physical case $N=2$
for a {\it fixed} atomic position $z = 1 \AA$ and a surface work-function
of $W = 4.59$ eV.  Three different initial conditions are studied.}
\label{fig7}
\end{figure}

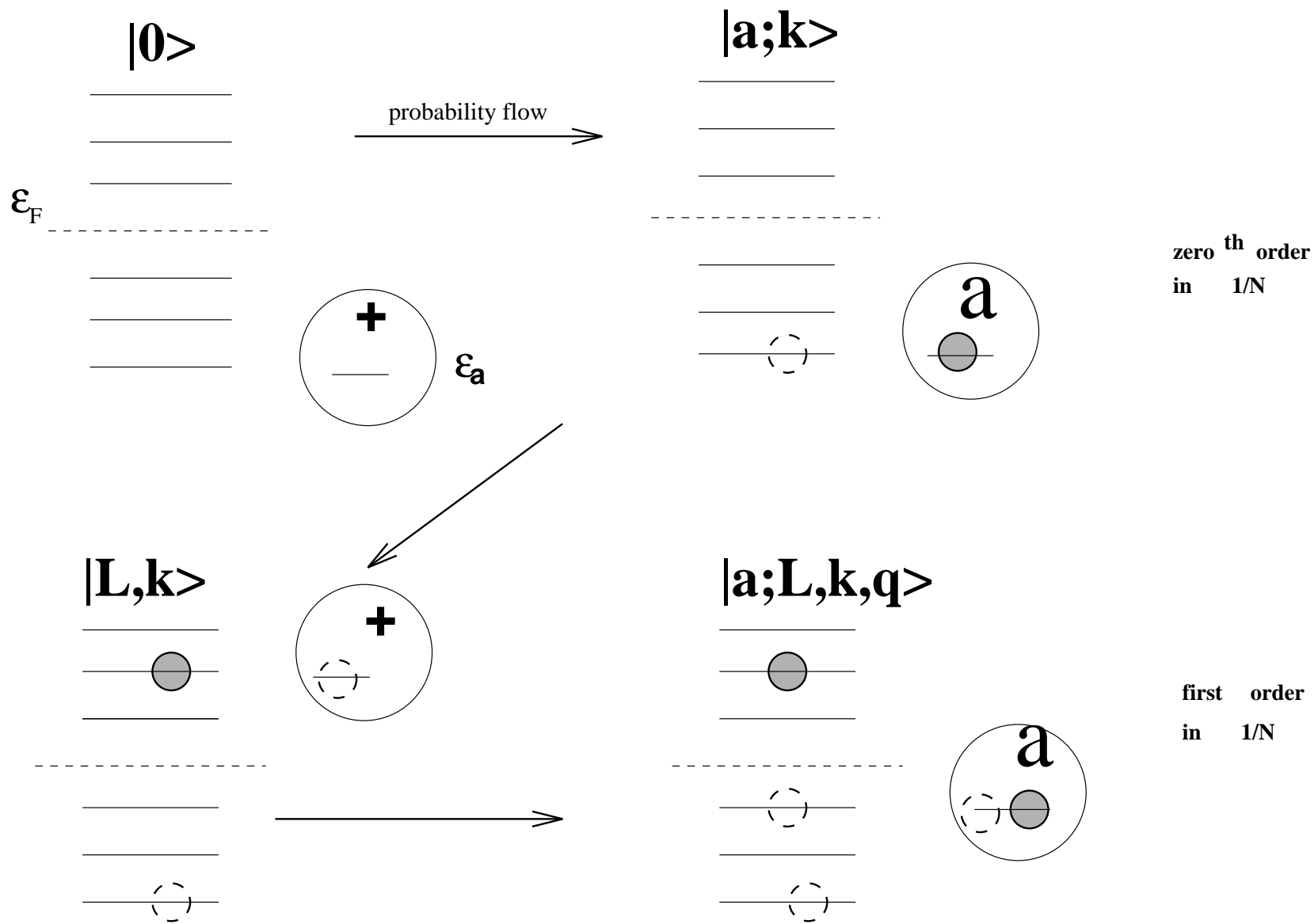
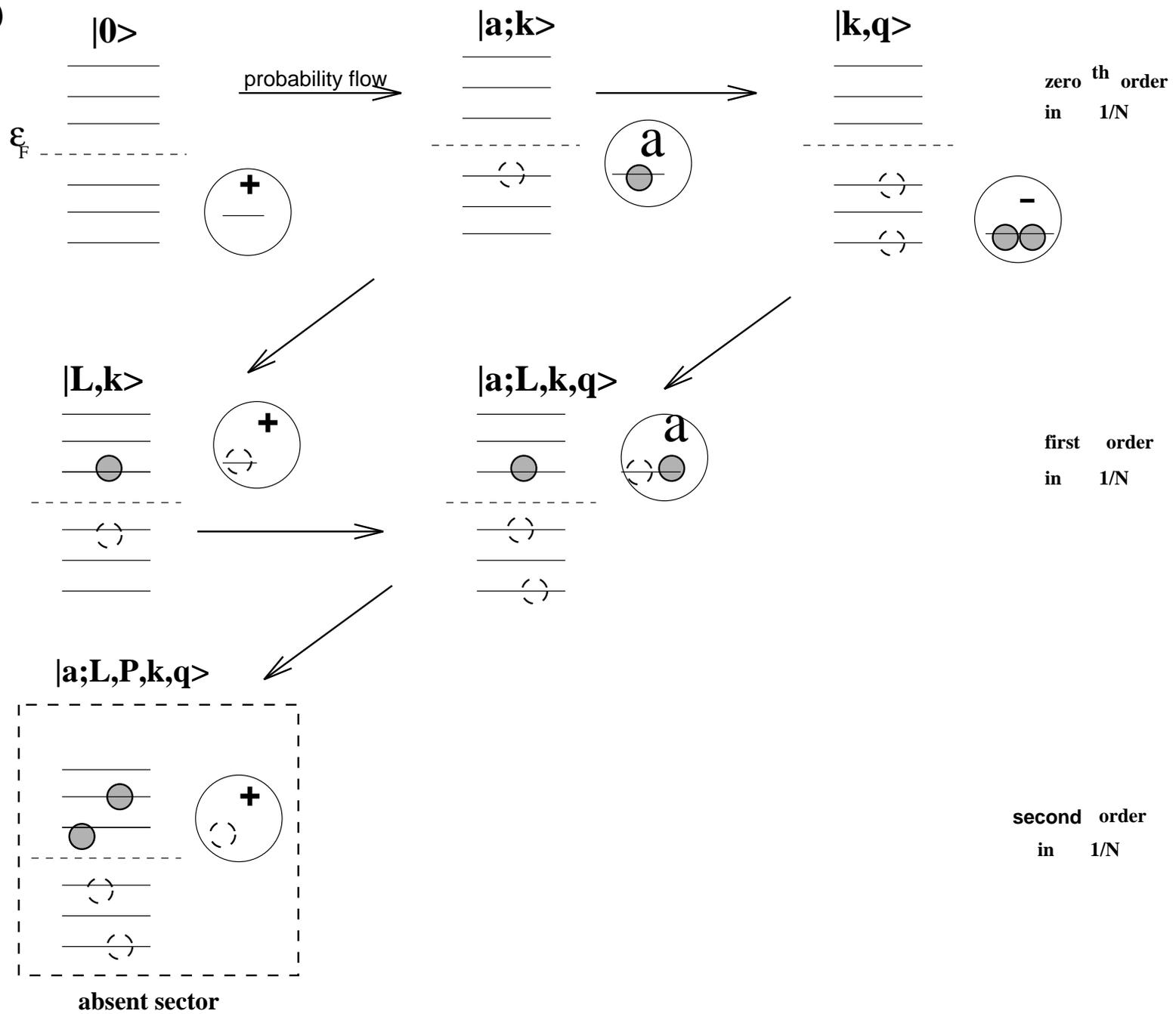
\begin{figure}
\caption{Schematic showing the different sectors of the Hilbert space
up to order ${1/N}$. The arrows indicate the direction
of probability flow as the entropy grows.
In (a), coupling to the negative ion $A^-$ is turned off and the corresponding
sector is not shown. In (b) all sectors discussed in
the paper are shown plus an $A^+$
sector at second order in $1/N$ which has not been
included.  We conjecture that loss-of-memory from an initial $A^-$ state
would occur if the Hilbert space were expanded further to include this sector.}
\label{fig8}
\end{figure}

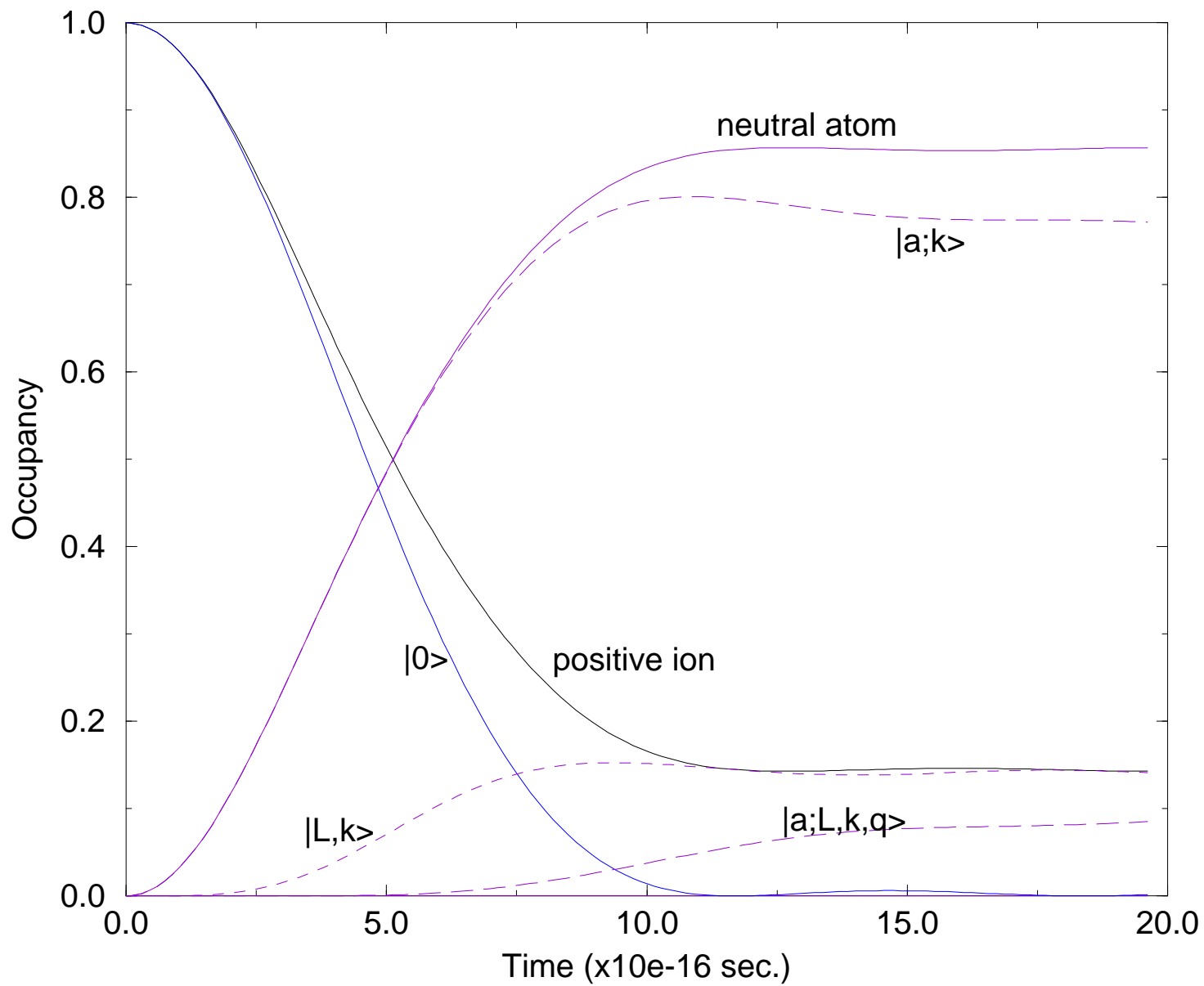
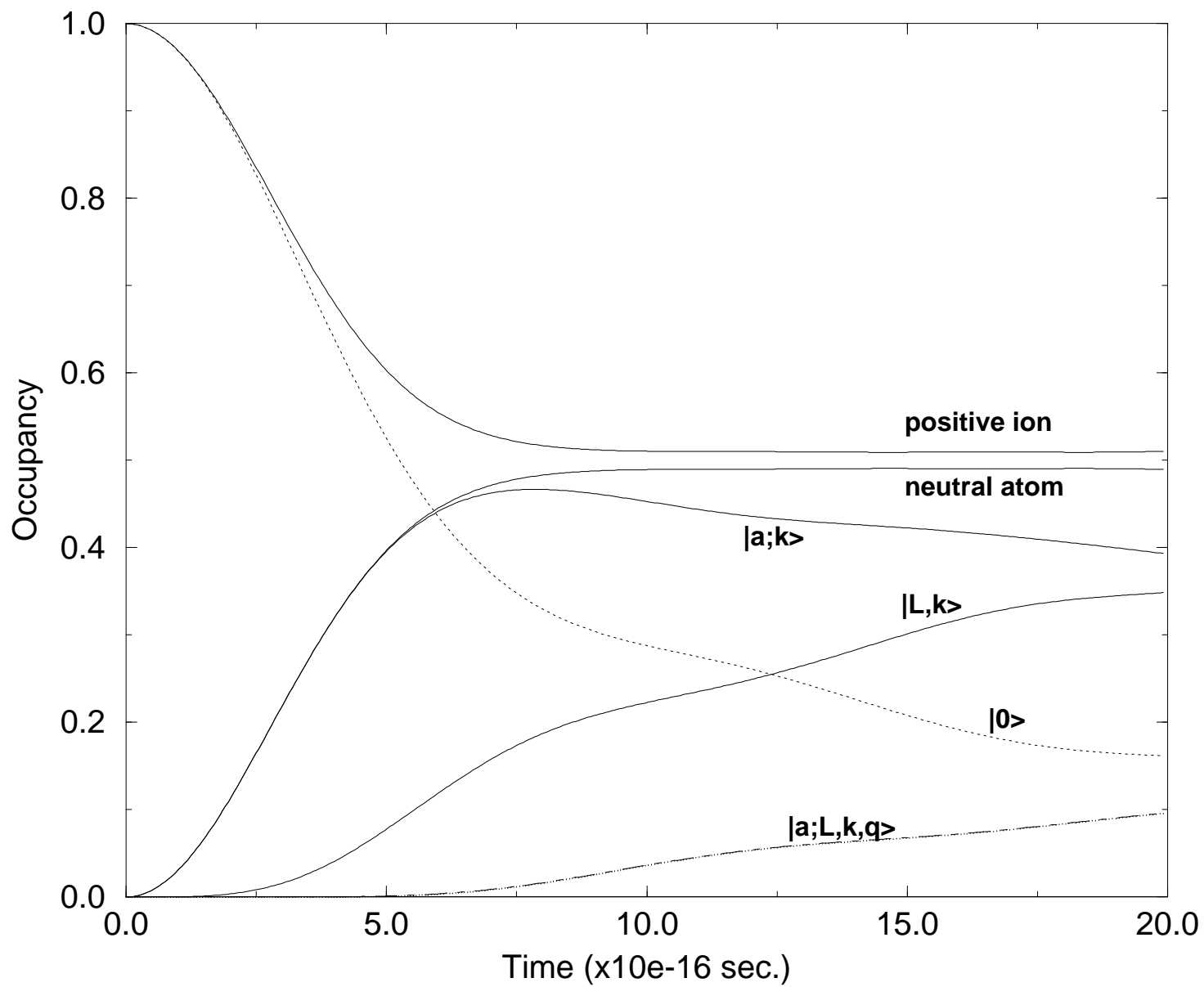
\begin{figure}
\caption{Occupancies as
a function of time for a {\it fixed} atomic position $z = 1 \AA$.  The
pure initial state is a positive ion (A$^+$) which then decays.  For clarity,
the coupling to the negative ion and excited neutral sectors is turned off.
(a) The surface work-function is $W = 1.59$ eV. (b) Same as (a) except the
surface work-function is $W = 3.28$ eV.  In both (a) and (b) a cascade of
probability flow from the low-order to the higher-order
sectors of the Hilbert space is evident.}
\label{fig9}
\end{figure}

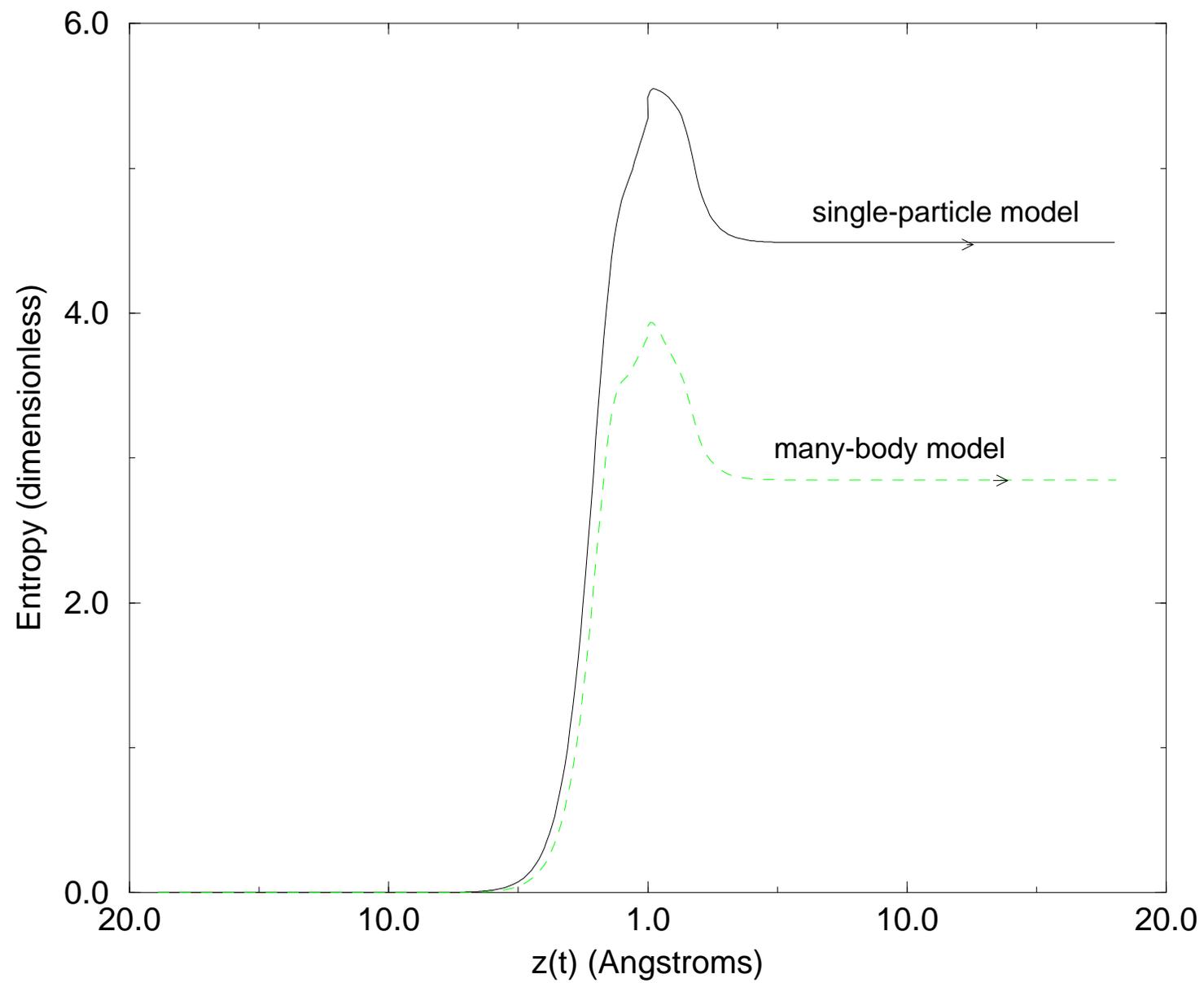
\begin{figure}
\caption{Time-evolution of the dimensionless entropy $S_{cg}[z(t)]$
in the full dynamical problem for spinless fermions ($N=1$).  We compare
the independent-particle solution with the many-body solution; only
the coupling to the Li(2s) state is turned on.
A positive Lithium ion with an incoming velocity of $u_i = 0.04$ a.u.
interacts with a metal surface of work-function $W=1.59$ eV.
The atom bounces off the surface with an outgoing velocity of $u_f = 0.03$ a.u.
and is completely neutralized. Note the comparable sizes of the two entropies.
In this case, the probability for an electron to be excited into a state
above the Fermi energy is only 0.098; hence the probability for the formation
of a particle-hole pair is also small.}

\label{fig10}
\end{figure}

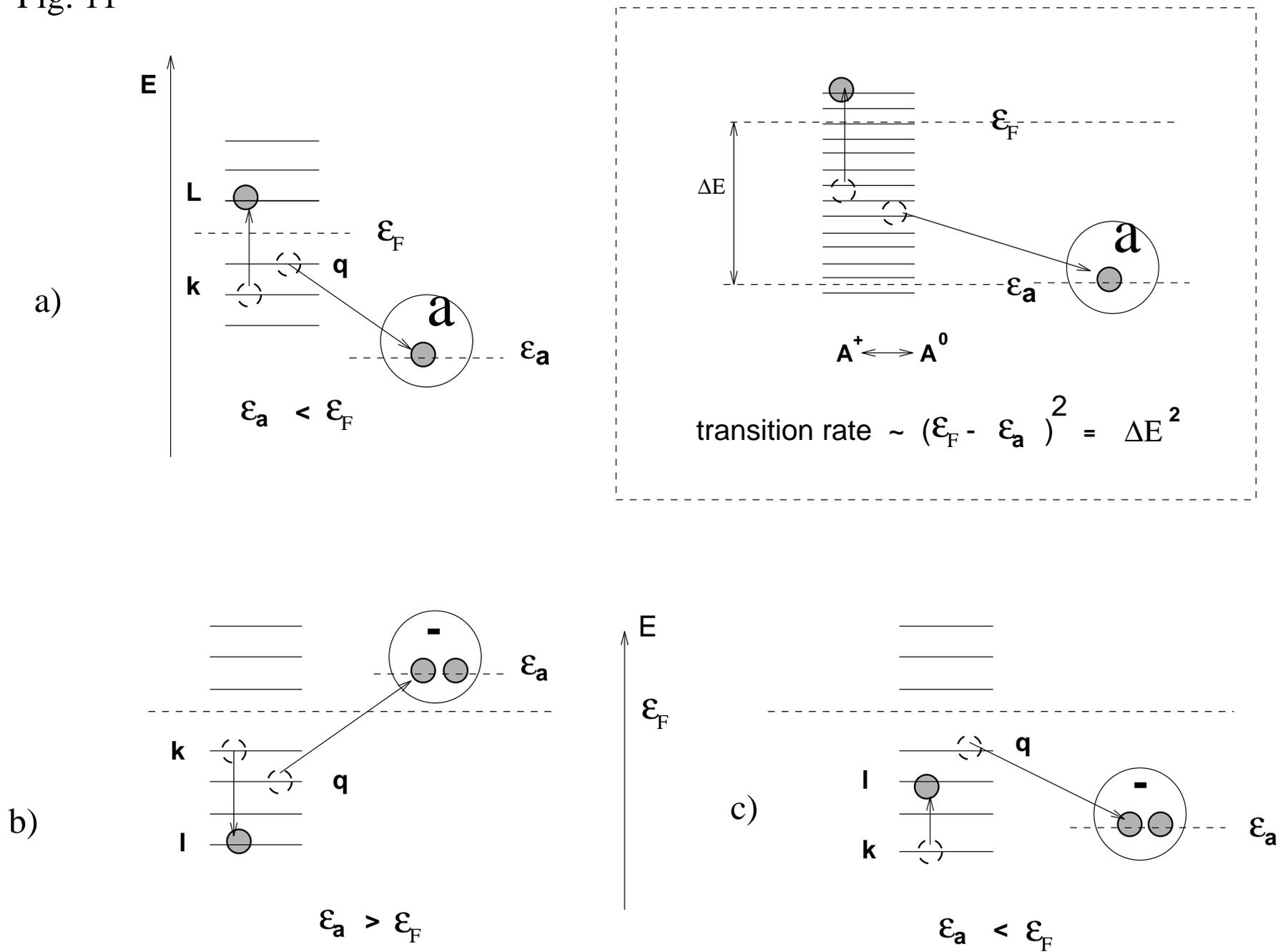
\begin{figure}
\caption{Schematic of three different Auger processes.
(a) A metal electron from below the Fermi level
of momentum $k$ transfers non-resonantly to
the atomic orbital $a$ while another electron from below the Fermi level
of momentum $q$ is promoted to a state of momentum $L$ above the Fermi level
to conserve energy.  Final state transitions are possible only when the
atomic level dips below the Fermi level, $\epsilon_{a} < \epsilon_{F}$.
(b) and (c) Two other Auger processes which involve the negative ion.
A metal electron from below the Fermi level transfers non-resonantly to
the atomic orbital $a$, while another electron from below the Fermi level of
momentum $k$ jumps to a state of momentum $l$ which is also below the Fermi
level.  Final state transitions in this case are possible both for
(b) $\epsilon_{a} > \epsilon_{F}$ and for
(c) $\epsilon_{a} < \epsilon_{F}$.}
\label{fig11}
\end{figure}

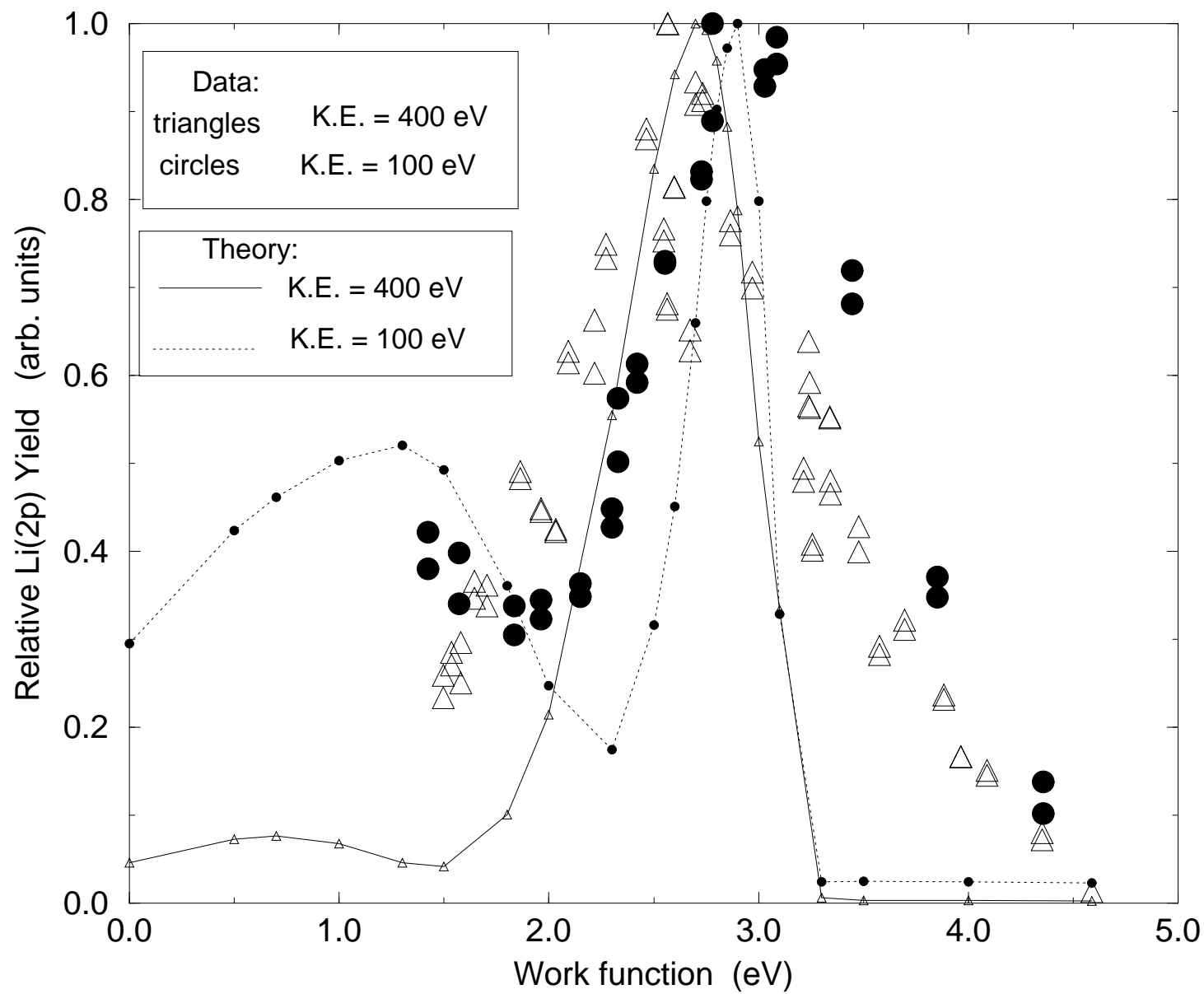
\begin{figure}
\caption{The experimentally observed and theoretically predicted
yield of excited neutral Li$^{0}(2p)$.  In the experiment, an incident
Li$^+$ ion
interacts with a metal surface of variable work-function.  In the theory,
$N=2$, $M = 30$,
and parameters which define the level widths due to Auger transitions
are given in Table II.  The initial condition in this case is (4),
the equilibrium ground state at point of closest approach, though similar
curves are also obtained for initial condition (2), the neutral atom far
away.  The solid line is for the case of $u_f = 0.03$ a.u. corresponding
to an incoming kinetic energy of $400$ eV.  The dashed line corresponds to
$u_f = 0.015$ a.u. or $100$ eV.  Yields are normalized to one.}
\label{fig12}
\end{figure}


\begin{references}

\bibitem{M1} J.B. Marston et.al., Phys. Rev. B {\bf 48}, 7809 (1993).

\bibitem{VY} C. M. Varma and Y. Yafet, Phys. Rev. B {\bf 13}, 2950 (1976).
See also:
O. Gunnarsson and  K. Sch{\"o}nhammer, Phys. Rev. B {\bf 28}, 4315 (1983).

\bibitem{BN2}R. Brako and D. M. Newns, Solid State Commun. {\bf 55}, 633
(1985).

\bibitem{BC} D. R. Andersson, E. R. Behringer, B. H. Cooper, and J. B. Marston,
{\it Journal of Vacuum Science and Technology} {\bf A11(4)},
2133 -- 2137 (1993); E. R. Behringer, D. R. Andersson, B. Kasemo, B. H. Cooper,
and J. B. Marston, {\it Nuclear Instruments and Methods in
Physics Research} {\bf B78}, 3 -- 10 (1993).

\bibitem{Weare}C. B. Weare and J. A. Yarmoff, ``Resonant Neutralization of
$^7$Li$^+$ Scattered from Alkali/Al(110) as a Probe of the Local
Electrostatic Potential,'' submitted to Surf. Sci.

\bibitem{Hermann} J. Hermann et al., Surf. Sci. {\bf138}, 570 (1984);
B. Hird {\it et al.,} Phys. Rev. Lett. {\bf 67}, 3375 (1991).

\bibitem{LG}J. Los and J. J. C. Geerlings, Phys. Reports {\bf 190}, 133 (1990).

\bibitem{Behringer} E.R Behringer, D.R. Andersson, B.H. Cooper and
J.B. Marston, unpublished.

\bibitem{HKM}H.-J. Kwon, A. Houghton, and J. B. Marston, Phys. Rev. B
{\bf 52}, 8002 (1995).

\bibitem{PN}P. Nordlander and J. C. Tully, Phys. Rev. Lett. {\bf 61},
990 (1988); Surf. Sci. {\bf 211/212}, 207 (1989); Phys. Rev. B {\bf 42},
5564 (1990).

\bibitem{PN2}P. Nordlander, Phys. Rev. B {\bf 46}, 2584 (1992).

\bibitem{Ernie} Ernest Behringer, {\it The Dynamics of Resonant
Charge Transfer In Hyperthermal Energy Ion-Surface Collisions}
(Ph.D. thesis, Cornell University, 1994)

\bibitem{BN1}R. Brako and D. M. Newns, Surf. Sci. {\bf 108}, 253 (1981).

\bibitem{Brown}To obtain a copy of the computer programs ``next.c'' and
``single.c'' contact J. B. Marston.

\bibitem{Shao2}Hongxiao Shao, Peter Nordlander, and David C. Langreth,
Phys. Rev. B {\bf 52}, 2488 (1995).

\bibitem{Shao}Peter Nordlander, Hongxiao Shao, and David C. Langreth,
Phys. Rev. B  {\bf 49}, 13929 (1994) and Phys. Rev. B {\bf 49}, 13948 (1994).

\bibitem{Hewson} A. C. Hewson, {\it The Kondo Problem to Heavy Fermions}
(Cambridge University Press, New York, 1993) p. 211.

\bibitem{Costi} T. A. Costi, J. Kroha, and P. W\"olfle, ``Spectral Properties
of the Anderson Impurity Model:  Comparison of Numerical Renormalization
Group and Non-Crossing Approximation,'' cond-mat/9507008, unpublished.

\bibitem{BN3}R. Brako and D. M. Newns, Rep. Prog. Phys. {\bf 52}, 655 (1989).

\bibitem{Dahl}E. Dahl, unpublished.

\bibitem{Davies} P.C.W. Davies, {\it The Physics of Time Asymmetry}
(University of California Press, Berkeley, 1974) pp. 157 -- 160.

\bibitem{Landau} L.D. Landau and E. Lifshitz, {\it Statistical Physics,
Part I} (Pergamon Press, Oxford, 1980).

\bibitem{Amos} A.T. Amos, B.L. Burrows, S.G. Davison, Surf. Sci. Lett.
{\bf 277}, 100 (1992).

\bibitem{Fonden}Tony Fonden and Andre Zwartkruis, Surf. Sci. {\bf 269/270},
601 (1992).

\bibitem{Negele} John W. Negele and Henri Orland, {\it Quantum Many-Particle
Systems} (Addison-Wesley Publishing Company, Redwood City, California, 1988)
p. 254.

\bibitem{Brenten}H. Schall, H. Brenten, K. H. Knorr, and V. Kempter,
Z. Phys. D {\bf 16}, 161 (1990); H. Brenten {\it et al.},Surf. Sci.
{\bf 243}, 309 (1991); Nucl. Instrum. Methods B {\bf 58}, 328 (1991).

\bibitem{Monreal}N. Lorente and R. Monreal, Nucl. Instrum. Methods
Phys. Res. B {\bf 78}, 44 (1993); R. Monreal and N. Lorente,
``Dynamical screening in Auger processes near metal surfaces,'' unpublished.

\end{references}
\end{document}